\documentclass[aps,prd,amsmath,amssymb,superscriptaddress,nofootinbib,10pt,showpacs,twocolumn]{revtex4-2}
\usepackage[]{latexsym}
\usepackage[english]{babel}
\usepackage{graphicx}
\usepackage{hyperref}
\usepackage{color}
\usepackage{bm}

\allowdisplaybreaks

\newcommand{\T}{\mathbf{\hat{\mathcal{T}}}}

\newcommand{\mI}{\mathcal{I}}

\newcommand{\mK}{\mathcal{K}}

\newcommand{\omegam}{\Omega_{\rm m}}

\newcommand{\Dk}[1]{\frac{d^3#1}{(2\pi)^3}}

\newcommand{\ve}[1]{{\text{\bf #1}}} 
\newcommand{\vk}{\ve k}

\newcommand{\vx}{\ve x}

\newcommand{\mA}{\mathcal{A}}
\newcommand{\mB}{\mathcal{B}}

\newcommand{\ikk}{\underset{\vk_{12}= \vk}{\int}}

\newcommand{\dD}{\delta_\text{D}}

\newcommand{\p}{\mathcal{L}}

\begin{document}
%opening
\title{On the galaxy 3-point correlation function in Modified Gravity}
\author{Alejandro Aviles}
\email{avilescervantes@gmail.com}
\affiliation{Departamento de F\'isica, Instituto Nacional de Investigaciones Nucleares,
Apartado Postal 18-1027, Col. Escand\'on, Ciudad de M\'exico,11801, M\'exico.}
\affiliation{Consejo Nacional de Ciencia y Tecnolog\'ia, Av. Insurgentes Sur 1582,
Colonia Cr\'edito Constructor, Del. Benito Ju\'arez, 03940, Ciudad de M\'exico, M\'exico}
\author{Gustavo Niz}
\email{g.niz@ugto.mx}
\affiliation{Departamento de F\'isica, Universidad de Guanajuato - DCI, 37150, Le\'on, Guanajuato, M\'exico.}

\pacs{PACS}

\begin{abstract}
The next generation of galaxy surveys will provide highly accurate measurements of the large-scale structure of the Universe, allowing for more stringent tests of gravity on cosmological scales.  Higher order statistics are a valuable tool  to study the non-Gaussianities in the matter field and to break degeneracies between modified gravity and other physical or nuisance parameters. However, understanding from first principles the behaviour of these correlations is essential to characterise deviations from General Relativity (GR), and the purpose of this work. This work uses contemporary ideas of Standard Perturbation Theory on biased tracers to characterize the three point correlation function (3PCF) at tree level for Modified Gravity models with a scale-dependent gravitational strength, and applies the theory to two specific models ($f(R)$ and DGP) that are representative for Chameleon and Vainshtein screening mechanisms. Additionally, we use a multipole decomposition, which apart from speeding up the algorithm to extract the signal from data, also helps to visualize and characterize GR deviations.

\end{abstract}

\maketitle

\begin{section}{Introduction}

One of the main goals in cosmology is to understand the nature of the gravitational interactions, which among other physical phenomena, are heavily responsible of the Universe's Large Scale Structure (LSS) distribution and evolution. Although the consistency of recent observations may point to physical tensions in the $\Lambda$CDM model \cite{Verde:2019ivm} and suggest deviations from the standard picture, the model based on General Relativity (GR) and a perturbative expansion about a homogeneous and isotropic background has proven extremely successful. However, the expansion history of $\Lambda$CDM may be easily reproduced by many Modified Gravity (MG) models, while it remains a challenge to mimic the growth of structure to all orders in perturbation theory. In this copycat approach of GR, some MG models may present an evolution of linear modes consistent with $\Lambda$CDM and present-day observations. Therefore, estimators that naturally test the nonlinear behaviour of perturbations may test gravity more efficiently. One of these estimators is the 3-Point Correlation Function (3PCF) or its Fourier space counterpart, the bispectrum, which encodes the shape of three-leg-interactions within the gravitational sector. Two point statistics are also sensitive to the non-linear gravitational evolution of perturbations, and in particular to three-leg interactions. However, their contribution is integrated over momentum loops so that a feature produced by a non-linear gravitational interaction at a particular scale may also be achieved by other physical phenomena. In contrast, three point statistics could break these degeneracies because, at leading order in perturbation theory, the operator is free on the momentum variable that is otherwise integrated in the two point statistics. As a result of this functional freedom, which leads to different triangular shapes, one has a richer structure worth studying. 

One can measure the three point statistics of density perturbations using a number of different physical observables. In particular, looking at matter tracers in the LSS may be a promising route to test gravity, due to the large number of modes and the high precision that near-future experiments, such as Dark Energy Spectroscopic Instrument (DESI) \cite{2016arXiv161100036D, 2016arXiv161100037D, 2022arXiv220510939A}, Euclid \cite{EUCLID:2011zbd} or the Vera Rubin Observatory \cite{LSST:2008ijt}, will achieve. In this paper, instead of focusing on baryonic tracers like  galaxies, clusters or the Lyman-alpha forest, we take one step back and study halos of two distinct but representative MG models. Although, our biasing scheme is general and our analytical results can be used for any kind of tracers. 

The theory space spanned by all possible extensions to GR is infinite. However, due to Lovelock's theorem \cite{Lovelock:1971yv,Lovelock:1972vz} one can categorise such extensions based on which of the theorem's assumption is not fulfilled. In particular, one assumption of the theorem is that gravity is mediated only by a helicity two field. Regarding this assumption, a popular choice to construct MG models is to  consider additional gravitational mediators to the metric. 
In here, we focus on using an extra scalar degree of freedom in the gravitational sector. Fifth force experiments on Earth and in the Solar System strongly restrict the nature of such additional fields (see for example \cite{Will:2014kxa}), so in order to obtain strong departures from GR on cosmological scales, one could impose a physical mechanism that screens the contribution of the extra degrees of freedom within the Solar System. There are three popular families of screening mechanisms (see \cite{Khoury:2010xi}): Chameleon \cite{Khoury:2003aq}, Symmetron \cite{Hinterbichler:2010es} and Vainshtein \cite{Vainshtein:1972sx}.  The models presented in this work are examples of the Chameleon and the Vainshtein mechanisms. In the Chameleon case, the mass of the scalar gravitational mediator depends on the local matter density, and representative models are the $f(R)$ theories (see \cite{DeFelice:2010aj} for an overview).  Here, we focus on one particular realisation called the Hu-Sawicky $f(R)$ model \cite{Hu:2007nk}, with different strengths of the modification. For the Vainshtein mechanism, the screening of the scalar field around matter sources is due to modified kinetic terms, and a canonical example, that we use in this work, is the DGP model \cite{Dvali:2000hr}. 
Although these models can resemble the expansion history of $\Lambda$CDM, the growth of structure would be modified, as have been studied for two point correlations. However, our purpose is to exhibit the rich structure of the three point statistics in the distribution of galaxies for these models. Moreover, differences between GR and these models in the 3PCF structure are not that degenerate among models or other physical contributions (such as the DM-galaxy bias parameters), supporting the idea of using this estimator to probe gravitational deviations from $\Lambda$CDM.

Calculating the 3PCF is a difficult task, given that it scales na\"{\i}vely as $N^3$, where $N$ is the number of galaxies in the catalogue. Using kd-trees or other common algorithms may reduce the computational time, but further developments are needed to obtain the 3PCF of millions of objects and on large scales. One recent approach to overtake this computational bottleneck is to use a multipole decomposition, reducing the algorithm to a roughly $N^2$ scaling, as shown by Slepian and Eisenstein \cite{Slepian:2015qza} (see also \cite{Slepian:2015qwa, Slepian:2017lpm}). This approach has been applied to obtain a 4.5$\sigma$ detection of Baryon Acoustic Oscillations (BAO) in the 3PCF \cite{Slepian:2016kfz} as well as the tightest current constraint on the way in which large-scale baryon–dark matter relative velocities couple to galaxy formation \cite{Slepian:2016nfb}. Both of these works used the CMASS sample of 777,202 Luminous Red Galaxies (LRGs) within the Sloan Digital Sky Survey (SDSS) Baryon Oscillation Spectroscopic Survey (BOSS). 

One can not only use this multipole expansion as a fast algorithm to extract the three point statistics signal from data, but also as a tool to visualise and classify differences between GR and the MG models. These differences in the 3PCFs manifest as a complex structure, which can be studied with two different approaches. On one hand, one could measure the 3PCF using the Slepian \& Eisenstein code on synthetic catalogues of modified gravity and GR. This approach has been carried on in \cite{Alam:2020jdv} over HOD catalogues obtained from the N-body set of simulations \verb|ELEPHANT|
\cite{Cautun:2017tkc} and using the same cosmological background for GR and MG. The HOD count-in-cell method was that of \cite{Hellwing:2017pmj}, while the evolution used the \verb|ECOSMOG| \cite{Li:2011vk,Bose:2016wms} and the \verb|ECOSMOG-V| \cite{Li:2013nua,Barreira:2015xvp} codes. On the other hand, there is a complementary approach to running 3PCF codes on synthetic data based on a theoretical description, which can explain from first principles the complex 3PCF structure found in \cite{Alam:2020jdv}. We devote this work to this theoretical approach, with particular attention to the multipole expansion of the 3PCF of modified gravity models, and employing Standard Perturbation Theory (SPT) at tree level. 

Before constructing the theory in detail, it is worth mentioning the efforts to describe MG models using perturbation theory. Most studies are dedicated to detail the non-linear matter power spectrum supplemented by diverse methods, such as the closure equations \cite{Koyama:2009me}, spherical collapse \cite{Brax:2013fna}, the multi-point propagator expansion \cite{Taruya:2014faa}, a peak-background split \cite{Bellini:2015oua}, semi-phenomenological treatments \cite{Fasiello:2017bot}, Lagrangian PT \cite{Aviles:2017aor}, EFT methods \cite{Bose:2018orj,Aviles:2020wme}, or the halo model \cite{Cataneo:2018cic}; and also extending the modeling to include redshift distortions \cite{Taruya:2013quf,Bose:2016qun,Aviles:2020wme} or biased tracers \cite{Aviles:2018saf}. Some of these PT prescriptions have been compared to N-body simulations to assess how well the theory captures the large and small scale clustering of biased tracers (see for example
\cite{,Valogiannis:2019xed}). 

Although most studies of the matter clustering in MG focus on two point statistics, there are studies which describe the matter bispectrum \cite{Hirano:2018uar,Bose:2018zpk,Bose:2019wuz}. However and to our knowledge, there is not any work on three point statistics in configuration space. Even less when considering the multipole basis expansion, which as discussed before allows for efficient 3PCF codes, that may be required to extract the 3PCF signal from the large datasets of the upcoming stage IV galaxy surveys. In this context, we revisit the PT framework of \cite{Slepian:2016weg}, and include the required ingredients to describe gravitational models beyond the standard one. 
The main ingredients are the inclusion of a new scale, introduced by the MG models, and the consideration of additional biasing operators, namely higher order biases, that are needed for theoretical consistency in 
MG (see e.g. \cite{Desjacques:2016bnm}). 

In summary and to describe the structure of the paper, we entrust ourselves to obtain a model of the 3PCF in a Legendre mulipolar basis for biased tracers. The route we take to this end is to Fourier transform the multipoles of the bispectrum; which is done in Sec.~\ref{subsec:Multipole3PCF}. However, before presenting the final results, in Sec.~\ref{subsec:MGmodels} we review the representative modified gravity models. In Sec.~\ref{subsec:SPTMG}, we review SPT on these MG models, following the ideas of Refs.~\cite{Koyama:2009me,Aviles:2017aor} up to second order in PT and with a biasing model, to finally construct the tree-level bispectrum for galaxies in Sec.~\ref{subsec:Multipole3PCF}. Finally, we include some conclusions in Sec.~\ref{subsec:concl}. Detailed derivations are delegated to appendices.

\end{section}

%%%%%%%%%%%%%%%%%%%%%%%%%%%%%%%%%%%%%%%%%%%%%%%%%%%%%%%%%%%%%%%%%%%%%%%%%%%%%%%%%
%%%%%%%%%%%%%%%%%%%%%%%%%%%%%%%%%%%%%%%%%%%%%%%%%%%%%%%%%%%%%%%%%%%%%%%%%%%%%%%%%

\begin{section}{Modified gravity theories with screening mechanisms}\label{subsec:MGmodels}

To study departures from General Relativity one has to choose a theoretical framework to characterise such deviations. Among the theories that violate the requirement of a metric as the gravity mediator in Lovelock's theorem \cite{Clifton:2011jh}, it is a popular choice to add an extra scalar gravitational degree of freedom.  If one restricts to second order differential equations and a minimally coupled matter sector, the Horndeski Lagrangian \cite{Horndeski:1974wa} is the most general scalar-tensor theory of gravity\footnote{Allowing for higher order equations of motion without introducing new degrees of freedom leads to the so called DHOST theories \cite{Langlois:2015cwa, Crisostomi:2016czh}}. Solar System constraints restrict the functional form of the unknown functions that define Horndeski's model, leaving a subset of options which contain screening mechanisms. A simple picture of screening mechanisms is achieved by looking at a conformally coupled scalar field to non-relativistic matter. As shown in \cite{Joyce:2014kja}, consider linear perturbations $\varphi$ of the scalar field around a background value $\varphi $, which in turn, is set by the local density of matter. Around a point-like mass source ($\rho=M\delta^3(r)$), the governing equation for the linear perturbation $\varphi$ can be schematically written as \cite{Joyce:2014kja}
\begin{align}\label{KG-all}
K(\varphi_0) \left[\ddot{\varphi}+c_s^2(\varphi_0)\nabla^2\varphi\right]+m^2(\varphi_0)\varphi=g(\varphi_0)M\delta^3(r).
\end{align}
For $K\sim g \sim c_s \sim 1 $ and $m\sim 0$, the additional gravitational force scales as $1/r^2$, violating all fifth force local constraints. However, by allowing $m$, $g$ and $K$ to depend on the environment, one may suppress the scalar interaction with matter altogether. The chameleon-family \cite{Khoury:2003aq} succeeds by providing a large mass to $\varphi$, leading to a rapidly decaying Yukawa effective potential, whereas the symmetron models \cite{Hinterbichler:2010es} decouple matter from the scalar degree of freedom via a weak coupling ($g\ll 1$). A third alternative to suppress the scalar interactions is to modify the kinetic term accordingly, through the function $K$. This mechanism, named after Vainshtein \cite{Vainshtein:1972sx}, can be constructed using either first (e.g. the K-mouflage \cite{Babichev:2009ee}) or second order derivatives (e.g. the Galileons \cite{Nicolis:2008in}) in the Lagrangian. 
For the present work, we focus on two representative cases of the Chameleon and Vainshtein families: the Hu-Sawicki $f(R)$ \cite{Hu:2007nk} and nDGP \cite{Dvali:2000hr} models respectively.  However, it is important to stress that the formalism presented here is more general than these two particular working examples. To set a common framework to discuss these MG models, we assume the same background expansion history for all models, governed by Friedmann's constraint $H^2=H_0^2(\Omega_{m}a^{-3}+\Omega_{\Lambda})$, where $a$ is the scale factor, $H=\dot{a}/a$ the Hubble growth rate,  and $\Omega_{m}$ and $\Omega_{\Lambda}$ are today's matter and dark energy abundances respectively. Dots refer to time derivatives and a zero subscript to quantities evaluated today (e.~g.~$a_0=a(t_0)=1$).

\subsubsection*{Hu-Sawicki $f(R)$ model}

The Hu-Sawicki (HS) model \cite{Hu:2007nk} is an $f(R)$ theory, where the $\Lambda$CDM Lagrangian, $R+f(R)=R-6H_0^2\Omega_{\Lambda}$, is replaced in the high curvature regime by the function $f(R)=-6H_0^2\Omega_{\Lambda}+ |f_{R0}|^n(R_0^2/R)^n$, where $R_0=3H_0^2(\Omega_{m}+4\Omega_{\Lambda})$ is today's Ricci scalar and $f_{R0}$ is the present day value of the additional scalar field gravitational mediator ($\varphi\sim f_R\equiv \partial f/\partial R$). For $n=1$, the weak field equations in the quasi-static limit for the Newtonian potential, $\Phi$, and the scalar field perturbation, $\varphi$, read \cite{Koyama:2009me}
%\begin{align}
%    \frac{1}{a^2}\nabla^2 \Phi(\vx) &= 4 \pi G \bar{\rho}\delta(\vx) - {1}{2a^2}\nabla^2 \varphi(\vx),  \label{poisson} \\
%    \frac{1}{a^2}\nabla^2 \varphi(\vx) &= -8 \pi G \bar{\rho} \beta^2 \delta (\vx)+ \sum_n \frac{1}{3n!}M_n\varphi^n(\vx), \label{KG}
%\end{align}
\begin{align}
    \frac{1}{a^2}\nabla^2 \Phi(\vx) &= 4 \pi G \bar{\rho}\delta(\vx) - \frac{1}{2a^2}\nabla^2 \varphi(\vx),  \label{poisson} \\
    \frac{1}{2 \beta^2 a^2} \nabla^2 \varphi(\vx) &= -8 \pi G \bar{\rho} \delta (\vx)+ \mI(\varphi,\nabla \varphi,\nabla\nabla \varphi,\dots), \label{KG}
\end{align}
where the interaction term $\mI$ is expanded in Fourier space as 
\begin{align} \label{mI}
\mI(\vk) &= M_1 \varphi(\vk) + \frac{1}{2}\int \frac{d^3\vk_1 d^3\vk_2}{(2\pi)^3} \dD(\vk-\vk_1-\vk_2) \nonumber\\ 
&\qquad \quad \times M_2(\vk_1,\vk_2) \varphi(\vk_1) \varphi(\vk_2) \, + \, \cdots.    
\end{align}
As long as the interaction depends only on the field $\varphi$ and not in its derivatives, the kernels $M_n$ are scale independent. This is the case of $f(R)$ theories, where
\begin{align}\label{Mn}
    M_n\equiv \frac{d^n R(\varphi)}{d\varphi^n}\big|_{\varphi=f_{R0}}, 
\end{align}
which are obtained by inverting $R$ as a function of $f_R$. Further, $\beta^2=1/3$ in these theories. Introducing the constant $\beta$ at this level is only for convenience, but will be relevant for the DGP model where it is a function of time. Notice that the associated mass in the previous Klein-Gordon equation for the additional scalar field depends on time, and it is given by \cite{Koyama:2009me}
%
%For $f(R)$ theories, the function $\beta(t)$ is $\beta^2=1/3$ and the $M_n$ are functions of time only given by
%\begin{align}\label{Mn}
%    M_n\equiv \frac{d^n R(\varphi)}{d\varphi^n}\big|_{\varphi=f_{R0}}, 
%\end{align}
%which are obtained by inverting $R$ as a function of $f_R$. Introducing the constant $\beta$ at this level is only for convenience, but will be relevant for the DGP model. Notice that the associated mass in the previous Klein-Gordon equation for the additional scalar field depends on time, and it is given by \cite{Koyama:2009me}
\begin{align}\label{massfR}
   m(a) \equiv \sqrt{2\beta^2 M_1} = \sqrt{\frac{M_1^\text{HS}(a)}{3}} \propto |f_{R0}|^{-1/2} ,
\end{align}
with
\begin{align} \label{M1fR}
M_1^\text{HS}(a) &= \frac{3 H_0^2 (\Omega_m a^{-3}+4\Omega_\Lambda)^3}{2 |f_{R0}|(\Omega_m +4\Omega_\Lambda)^2}; %\quad
% M_2^{f_R}(a) = \frac{9 H_0^2 (\Omega_m a^{-3}+4\Omega_\Lambda)^5}{4 |f_{R0}|^2(\Omega_m a^{-3}+4\Omega_\Lambda)^4},   
\end{align}
while the first auto-interaction is mediated by the dynamical coupling \cite{Koyama:2009me}
\begin{align}
% M_1^{f_R}(a) &= \frac{3 H_0^2 (\Omega_m a^{-3}+4\Omega_\Lambda)^3}{2 |f_{R0}|(\Omega_m a^{-3}+4\Omega_\Lambda)^2}, \quad
M_2^\text{HS}(a) = \frac{9 H_0^2 (\Omega_m a^{-3}+4\Omega_\Lambda)^5}{4 |f_{R0}|^2(\Omega_m +4\Omega_\Lambda)^4}.   
\end{align}
Therefore, the theory results in a Yukawa-like interaction at leading order for the fifth force, with further corrections due to the non-linear interactions. The associated Compton scale to the Yukawa-like force is $m^{-1}\propto |f_{R0}|^{1/2}$, which for larger values of $|f_{R0}|$ results in larger deviations from GR. In the present work and to exemplify our modeling numerically, we assume two possible strengths of $f_{R0}$: $\text{F4}\sim 10^{-4}$ and $\text{F6}\sim 10^{-6}$. 

\subsubsection*{nDGP model}
The Dvali, Gabadadze and Porrati (DGP) model is based on a four dimensional braneworld which contains the matter fields, embedded in a five dimensional spacetime. There is cross over scale, $r_c$, so that below it the model is effectively 4d GR plus a scalar field which captures the flexing or bending of the brane. In this $r<r_c$ regime, the Poisson and Klein-Gordon (KG) equations in the quasi-static limit can be cast in the same way as for the $f(R)$ model, namely Eqs.~(\ref{poisson})-(\ref{KG}), but with the interaction $\mI$ depending on the field second derivatives as
\begin{equation}
    \mI = \frac{r_c^2}{a^2} \left[ (\nabla^2 \varphi)^2 -(\nabla_i \nabla_j \varphi)^2 \right],
\end{equation}
and hence the only surviving kernel $M_n$ is
\begin{align}  \label{M2DGP}
M_2^\text{DGP}(\vk_1,\vk_2)&= \frac{2 r_c^2}{a^4}\big[k_1^2k_2^2-(\vk_1\cdot\vk_2)^2\big]. 
\end{align}
%Further, since $M^\text{DGP}_{1}=0$, the mass vanishes.
%\gn{Is there a beta in M2?} 
The time-dependent coupling $\beta$ is 
\begin{align}\label{betaDGP}
  \beta^2(t) = \frac{1}{6}\left[1+2H r_c\left(1+\frac{\dot{H}}{3H^2} \right) \right]^{-1}.
\end{align}
Notice, in the DGP literature it is common to find a different definition for $\beta$, where $\beta^2$ in this work is $(6\beta)^{-1}$ in other works. Models with $r_c= \text{X}/H_0$, with X a positive number are denoted as NX. In this work we will focus on the N1 and N5 {\it normal branch} DGP models.

%\begin{subsection}{N-body simulations and halo catalogues}

%\end{subsection}

%%%%%%%%%%%%%%%%%%%%%%%%%%%%%%%%%%%%%%%%%%%%%%%%%%%%%%%%%%%%%%%%%%%%%%%%%%%%%%%%%
%%%%%%%%%%%%%%%%%%%%%%%%%%%%%%%%%%%%%%%%%%%%%%%%%%%%%%%%%%%%%%%%%%%%%%%%%%%%%%%%%

So far, we have introduced all the needed basics of the HS and DGP representative MG models in order to understand how to get the tree-level bispectrum using SPT, which is described in the next section.

\begin{section}{Standard Perturbation Theory in Modified Gravity}\label{subsec:SPTMG}

The MG theories from the previous section belong to a class of theories where the linear growth function $D_+$ is the fastest-growing solution of the differential equation\footnote{Notice at linear order eqs.~\eqref{poisson}, \eqref{KG} and \eqref{mI} imply $(k^2  /a^2) \Phi=- A(k) \delta$; see also Appendix \ref{app:F2}.}
\begin{align}\label{DplusEq}
\left[\T - A(k,t) \right] D_+(k,t) = 0, 
\end{align}
with the linear differential operator $ \T = \partial^2/\partial t^2 + 2 H (\partial/\partial t)$ \cite{Matsubara:2015ipa},
and a scale-dependent gravitational strength defined as
\begin{align} \label{defAk}
 A(k,t) = \frac{3}{2} \omegam H^2 \left(1 +\frac{2 \beta^2 k^2}{k^2 + m^2 a^2} \right). 
 \end{align}
The quantities $m(a)$ and $\beta(a)$ are model dependent functions, 
that respectively quantify the range and the strength of the fifth force; $m(a)$ is typically the associated effective mass of the new scalar gravitational degree of freedom, and $\beta$ its coupling to the other matter fields, which is assumed universal. As long as $m$ is different from zero, the linear growth function $D_+$ is scale dependent and the theories reduce to GR at large scales  $k\ll m a$; 
this is the case of chameleons, as the Hu-Sawicky $f(R)$ model previously discussed; see Eq.~\eqref{M1fR}. %Actually in this $f(R)$ model, the function $\beta^2=1/3$ is constant, as appreciated from (\ref{fReoms}), while the mass $m$ evolves with time, as previously discussed in Eq.~(\ref{M1fR}).

In contrast, other theories such as DGP and cubic Galileons have zero mass, since for these $M_1=0$; thus the linear growth depends only on time. These models do not reduce to GR at large scales, implying they are tightly constrained by the background evolution. However, one can always add a smooth dark energy component to mimic as much as desired the $\Lambda$CDM expansion history \cite{Schmidt:2009sv}. This is usually done in $N$-body simulations in order to
isolate the effects of the growth of perturbations due to an extra force component from those effects due to a different overall background expansion history of the Universe.
Nevertheless, in such theories, additional scale dependencies enter beyond linear order through derivative couplings in their associated Klein-Gordon equations, as can be seen from Eq.~(\ref{KG}) with the lowest order correction set by $M_2$ of Eq.~(\ref{M2DGP}). %The lowest order correction is set by $M_2^{DGP}$, given by
This term will become relevant at second order in perturbation theory as we will discuss later. 

The choice of  Eq.~\eqref{defAk} may be seen as too restrictive for the HS or DGP (or cubic Galileon) models; however, in \cite{Bose:2016qun} it is shown that a large subset of the Horndeski sector can be written in this form at linear order, while non-linearities are correctly modeled by the $M_i(k,a)$ functions. Furthermore,  theories posed in the Einstein frame, such as symmetrons \cite{Pietroni:2005pv,Olive:2007aj,Hinterbichler:2010es}, can be easily recast in this form as well, by using field redefinitions \cite{Aviles:2018qot}, which are not conformal transformations of the metric tensor.\footnote{For the purposes of late time LSS formation, the main difference is that in Einstein frame theories the new scalar degree of freedom does not couple to the Poisson equation, as in Eq.~\eqref{poisson}, but to the Geodesic equation since particles follow geodesics of a conformally transformed metric.}

The linear matter overdensity at time $t$ is $\delta^{(1)}(\vk,t) = \big[ D_+(k,t)/D_+(k,t_0) \big] \delta^{(1)}(\vk,t_0)$,  with $t_0$ an arbitrary time, usually chosen to be the the present time.  When possible, it is convenient to normalize the linear growth function to Einstein-de Sitter (EdS) evolution, $D_+(k,t_{ini}) =  D_+^\text{EdS}(t_{ini}) \propto a(t_{ini})$, for some early initial time $t_{ini}$ where the linear growth evolution is indistinguishable to that in an EdS universe. This is possible for the majority of MG theories studied in the cosmological literature, where one can construct the linear power spectrum (PS) in terms of the $\Lambda$CDM linear PS as
\begin{align}
P_L(k,t) = \left[\frac{D_+(k,t)}{D_+^\text{$\Lambda$CDM}(k,t_0)}\right]^2 P_L^\text{$\Lambda$CDM}(k,t_0).
\end{align}
An alternative is to obtain the linear PS directly from an Einstein-Boltzmann code, such as \verb|MGCAMB| \cite{Lewis:1999bs,Hojjati:2011ix} or \verb|hi_class| \cite{Blas:2011rf,Zumalacarregui:2016pph}. 

To include higher order corrections in the correlation function, one may solve the hydrodynamic equations iteratively. To second order in the matter fluctuation one finds
\begin{align} \label{delta2}
\delta^{(2)}(\vk,t) &= \int \frac{d^3 k_1 d^3 k_2}{(2 \pi)^3} \dD(\vk - \vk_1 -\vk_2) F_2(\vk_1,\vk_2,t) \nonumber\\
                    &\quad \times\delta^{(1)}(\vk_1,t)\delta^{(1)}(\vk_2,t),
\end{align}
where the second order SPT kernel $F_2$ is given by (see Appendix \ref{app:F2} for a derivation)
\begin{align}\label{F2kernel}
 F_2(\vk_1,\vk_2,t) &= \frac{1}{2} + \frac{3}{14}\mA(\vk_1,\vk_2,t) + \frac{x}{2}\left[ \frac{k_1}{k_2} + \frac{k_2}{k_1}\right]\nonumber\\ 
                    &\quad  + x^2 \left[ \frac{1}{2} - \frac{3}{14}\mB(\vk_1,\vk_2,t) \right]  
 ,
\end{align}
with $x=\hat{\vk}_1 \cdot \hat{\vk}_2$ the angle 
between the two interacting density fields with wave-vectors $\vk_1$ and $\vk_2$.% in Eq.~(\ref{delta2}). 
The scale and time dependent functions 
$\mA$ and $\mB$ are set by
\begin{align} \label{AandBdef}
 \mA(\vk_1,\vk_2,t) &= \frac{7 D^{(2)}_{\mA}(\vk_1,\vk_2,t)}{3 D_{+}(k_1,t)D_{+}(k_2,t)}, \nonumber\\
 \mB(\vk_1,\vk_2,t) &= \frac{7 D^{(2)}_{\mB}(\vk_1,\vk_2,t)}{3 D_{+}(k_1,t)D_{+}(k_2,t)},
\end{align}
where we have defined a second order growth functions $D^{(2)}_{\mA,\mB}$, which are solutions of the Green's problem \cite{Aviles:2017aor}% (see Appendix \ref{app:F2} for a derivation)
\begin{align}
D^{(2)}_{\mA} &= \big(\T - A(k)\big)^{-1}\Bigg[A(k)   + (A(k)-A(k_1))\frac{\vk_1\cdot\vk_2}{k_2^2} \nonumber\\
%                            &     \qquad       +  \frac{2 \beta^2 k^2}{k^2 + m^2 a^2}\mK^{(2)}_{\delta I} \Big]  D_{+}(k_1)D_{+}(k_2), \label{DAeveq} \\
             &\quad   + (A(k)-A(k_2))\frac{\vk_1\cdot\vk_2}{k_1^2} 
             \nonumber\\
             &\quad  - S_2(\vk_1,\vk_2) \Bigg]  D_{+}(k_1)D_{+}(k_2), \label{DAeveq} \\
 D^{(2)}_{\mB} &= \big(\T - A(k)\big)^{-1}\Big[A(k_1) + A(k_2) - A(k) \Big] \nonumber\\
             &\quad \times D_{+}(k_1)D_{+}(k_2). \label{DBeveq}
\end{align}
In the previous expressions, the wavenumber $\vk = \vk_1 + \vk_2$, which follows from momentum conservation, mathematically expressed by the Dirac delta function in Eq.~(\ref{delta2}). 
%In solving the above equations, appropriate initial conditions should be chosen in order to project out the first order solution to Eq.~(\ref{DplusEq}). 
% The norm of the wave-vector $\vk$  entering Eqs.~(\ref{DAeveq,DBeveq}) is given by $k=|\vk_1+\vk_2|$.
An EdS background evolution
results in $\mA^{\text{EdS}} = \mB^{\text{EdS}} =1$; whereas for the $\Lambda$CDM model, $\mA(t) = \mB(t)$ are only weakly dependent on time and close to one. Actually,  
for standard cosmologies one finds that nowadays $\mA^{\Lambda\text{CDM}}(t_0) \simeq 1.01$.
However, in more general cases on which additional scales enter the theory, as in MG or massive neutrinos, the functions $\mA$ and $\mB$ are unequal and scale dependent. 
Hereafter, we omit to write the time dependencies of these and other related functions to simplify the expressions.

The source $S_2(\vk_1,\vk_2)$ in Eq.~(\ref{DAeveq}) comes from the non-linearities of the Klein-Gordon equation (\ref{KG}) and is responsible for the screening mechanism to second order in perturbation theory.
One finds, explicitly, that $S_2$ for the HS $f(R)$ model is
\begin{align}\label{sourceS}
S^\text{HS}_2(\vk_1,\vk_2) =    \frac{36 \Omega_m^2 H^4 \beta^6 a^4 M_2(\vk_1,\vk_2)k^2}{ (k^2 + m^2a^2)(k_1^2 + m^2a^2)(k_2^2 + m^2a^2)},
\end{align} 
where a more detailed derivation can be found in Appendix \ref{app:F2}.
% For the HS, $f(R)$ model, the function $\beta^2=1/3$ is constant  while the effective mass is \cite{Koyama:2009me}
% \begin{align}\label{massfR}
%    m(a)=\sqrt{\frac{M_1(a)}{3}} \propto |f_{R0}|^{-1/2} 
% \end{align}
% with
% \begin{align} \label{M1fR}
% M_1^{f_R}(a) &= \frac{3 H_0^2 (\Omega_m a^{-3}+4\Omega_\Lambda)^3}{2 |f_{R0}|(\Omega_m +4\Omega_\Lambda)^2}. %\quad
% % M_2^{f_R}(a) = \frac{9 H_0^2 (\Omega_m a^{-3}+4\Omega_\Lambda)^5}{4 |f_{R0}|^2(\Omega_m a^{-3}+4\Omega_\Lambda)^4},   
% \end{align}
% While function $M_2$ is
% \begin{align}
% % M_1^{f_R}(a) &= \frac{3 H_0^2 (\Omega_m a^{-3}+4\Omega_\Lambda)^3}{2 |f_{R0}|(\Omega_m a^{-3}+4\Omega_\Lambda)^2}, \quad
% M_2^{f_R}(a) = \frac{9 H_0^2 (\Omega_m a^{-3}+4\Omega_\Lambda)^5}{4 |f_{R0}|^2(\Omega_m +4\Omega_\Lambda)^4}.   
% \end{align}
% Functions $M_{i}^{f_R}$ are only time dependent for $f(R)$ theories.
% For DGP, given that it is massless one has $M_1^\text{DGP}=0$. Furthermore,
% \begin{align} \label{M2DGP}
% M_2^\text{DGP}(\vk_1,\vk_2)= \frac{2 r_c^2}{a^4}\big[k_1^2k_2^2-(\vk_1\cdot\vk_2)^2\big].
% \end{align}
If one focuses on the DGP model, the leading correction comes from $M_2^\text{DGP}$, because as discussed before this model has a vanishing mass. An immediate consequence of the explicit form of $M_2^\text{DGP}$ is that the screening source term depends only on the angle $x$ between the two interacting plane waves. To appreciate this result, observe that Eqs.~\eqref{S2def} and \eqref{M2DGP} reduce to
\begin{align}\label{sourceSDGP}
 S^{\text{DGP}}_2(x) =  Z(t) (1-x^2),
\end{align}
with 
\begin{align}
    Z(t) = 72 \,\Omega_m^2 H^4 \beta^6 r_c^2.
\end{align}
A direct consequence of this result is that the second order growth function $D^{(2)}_\text{DGP}(\vk_1,\vk_2)$ depends on the form of the triangle formed by the wavevectors 
$\vk=\vk_1+\vk_2$, $\vk_1$ and $\vk_2$, but not on its size. Notice that this scaleless dependence comes only from the screening terms, as discussed in an extended manner in \cite{Aviles:2018qot}.  
At higher than second order this is no longer true, because in Eq.~\eqref{mI} we are expanding $\delta \mI$ in terms of (non-linear) fields $\varphi$, which in turn should be expanded into linear densities, and when properly done, Eq.~\eqref{mI} receives contributions at all orders in PT regardless of $M^\text{DGP}_{n>2}=0$. However, these appear for the first time at 1-loop level, and at leading order congruent triangles have identical screenings.

For the present work purpose, it is sufficient to consider up to second order in the density fields in order to obtain the matter tree-level bispectrum, whose computation follows the usual expression
\begin{align}\label{matterBisp}
 B(\vk_1,\vk_2,\vk_3) &= 2F_2(\vk_1,\vk_2) P_L(k_1) P_L(k_2)  + \text{\, cyclic}, 
\end{align}
where the last term represents the cyclic permutations of the wave-vectors $\vk_1$, $\vk_2$, $\vk_3$.   

%\alej{Perhaps write here the 3PCF for matter} 

\end{section}

\begin{section}{Tree-level Bispectrum of Galaxies}\label{subsec:treeB}

We are interested in the statistics of galaxies as biased tracers of the underlying dark matter distribution. The standard biasing approach consists on constructing all the relevant operators consistent with symmetries up to the desired order in PT \cite{McDonald:2009dh,Assassi:2014fva}. In $\Lambda$CDM with initially Gaussian distributed matter fields, it is sufficient to  consider the local\footnote{By {\it local} we mean {\it local-in-matter-density}, as it is used in the more recent literature; see e.g. \cite{Desjacques:2016bnm}.}  operators $\delta$ and $\delta^2$, and the tidal field $s^2$ for the leading order bispectrum.
However, the completeness of this set of operators relies on the fact that the linear growth functions are scale independent, such that all modes grow at the same pace, which is not true for MG models in general. In fact, it is well known that even the linear local bias is scale dependent in MG, scaling as $b_1(k,t) = 1 + D_+(k,t_*)/D_+(k,t)(b_1(k,t_*) - 1)$ \cite{Hui:2007zh,Parfrey:2010uy}, hence becoming non-multiplicative in configuration space, unless $D_+$ is separable in $k$ and $t$ and depends on time only at some sufficiently early time. We note, however, that for larger scales than the scalar field fifth-force range ($\sim 1/m$), we can expand the gravitational strength $A(k)$ in powers of $(k/am)^2$, and by considering operators $\nabla^2 \delta$, $\nabla^4 \delta$, and so on (commonly named higher-derivative or curvature operators) we deal effectively with the effects of MG at sufficiently large scales. This approach was taken in  \cite{Aviles:2018saf,Aviles:2020wme},  following the discussion of Sec.~8.3 in ref.~\cite{Desjacques:2016bnm}. 
Moreover, curvature bias is also well motivated in $\Lambda$CDM by the formation of halos in peaks theory \cite{Kaiser:1984sw,Bardeen:1985tr,Desjacques:2010gz,Lazeyras:2015giz}. Further, it is often used to remove subleading dependencies on the smoothing scale $R$ of the matter overdensity $\delta_R$ (that we simply write here as $\delta$ for compactness) \cite{McDonald:2009dh,Schmidt:2012ys,Aviles:2018thp}. Hence, our biasing model use the tracers density fluctuation expanded as

\begin{align} \label{halodensity}
%  \delta_h(\vx) = b_0 + b_1\delta(\vx) + \frac{b_2}{2} \delta^2(\vx) + \frac{1}{2} b_{s^2} (s_{ij})^2 + \cdots    %1407.5668  Gil-Marin 2015
 \delta_g(\vx) &= b_0 + b_1\delta(\vx) + b_{\nabla^2\delta} \nabla^2 \delta(\vx) \nonumber\\
 &\quad + \frac{b_2}{2} \delta^2(\vx) + b_{s^2} s^2(\vx) + \cdots , 
\end{align}
with $s^2 = s_{ij} s^{ij}$ the tidal bias operator and
\begin{align}
 s_{ij}(\vx) = \left(\frac{\partial_i \partial_j}{\nabla^2} - \frac{1}{3}\delta_{ij}\right) \delta(\vx),
\end{align}
the shear tensor. The parameter $b_0$ contains all term necessary for canceling out zero-lag correlators: that is, $b_0 = -\frac{b_2}{2} \langle \delta^2 \rangle - b_s^2 \langle s^2 \rangle$ up to second order in PT and bias expansion.\footnote{In Ref.~\cite{Slepian:2016weg}, the authors use the notation $b_t = 2 b_{s^2}$ for the tidal bias. Also, the second order local bias in that work is related to ours by $2 b_2^\text{[that work]} = b_2^\text{[here]}$. However, for the definitions of $\gamma$ and $\gamma'$, introduced below, both notations coincide.} 
% Note in ref.~\cite{Slepian:2016weg} (hereafter SL16) the authors use $b_t = 2 b_{s^2}$ for the tidal bias. 
Bias operators $\delta \nabla^2 \delta$, $\nabla \delta \cdot \nabla \delta$ and $(\nabla^2 \delta)^2$ can also be considered at second order, but they yield very small contributions at large scales, thus we neglect them in this work.

Additionally to the biasing expansion, we use the PT formal expansion of the galaxy density field $\delta_g(\vx) = \delta^{(1)}_g(\vx) + \delta^{(2)}_g(\vx) + \cdots$. The double expansion in fluctuations and biases in Fourier space results in the first and second order galaxy fields
\begin{align}
 &\delta^{(1)}_g(\vk) = \left(b_1 - b_{\nabla^2\delta} k^2  \right)\delta^{(1)}(\vk),  \\
 &\delta^{(2)}_g(\vk) = \int \frac{d^3 k_1 d^3 k_2}{(2 \pi)^3} \dD(\vk-\vk_1-\vk_2) \nonumber\\ 
 &\quad \times \Big[ \left(b_1 - b_{\nabla^2\delta} k^2 \right) F_2(\vk_1,\vk_2)  + \frac{b_2}{2} \nonumber\\
 &\qquad  + b_{s^2} \left((\hat{\vk}_1 \cdot \hat{\vk}_2)^2 - \frac{1}{3}\right) \Big] \delta^{(1)}(\vk_1)\delta^{(1)}(\vk_2),
\end{align}
with the second order SPT kernel $F_2$ given by Eq.~(\ref{F2kernel}).
 % The scale $a^2 m^2(a)$ appears from expanding the Klein-Gordon equation in powers of $k^2/a^2 m^2(a)$. 
 
Our main object of study is the galaxy-galaxy-galaxy bispectrum, $B$, that at tree-level is defined through 
\begin{align}
 &\langle \delta^{(1)}_g(\vk_1)\delta^{(1)}_g(\vk_2)\delta^{(2)}_g(\vk_3)\rangle \nonumber\\
 &\quad  = (2\pi)^3 \dD(\vk_1+ \vk_2+ \vk_3)B(\vk_1,\vk_2,\vk_3)  + \text{cyclic},
\end{align}
%with ``cyclic'' denoting two cyclic permutations of wavevectors $\vk_1$, $\vk_2$ and $\vk_3$, as in the equation for the matter bispectrum (\ref{matterBisp}).
which after some manipulations becomes %By developing the above equations we obtain
\begin{widetext}
\begin{align} \label{bispectrum}
 B(\vk_1,\vk_2,\vk_3) &= b_1^3 \left(1+ \gamma_* k^2_1 \right) \left(1 + \gamma_* k^2_2 \right) P_L(k_1)P_L(k_2) 
 %\nonumber\\ & \times
 \left[ 2 F_2(\vk_1,\vk_2) \left(1 + \gamma_*  k^2_3 \right) + \gamma + 2\gamma' \left((\hat{\vk}_1 \cdot \hat{\vk}_2)^2 - \frac{1}{3}\right)\right] 
 \nonumber\\ &\quad
 + \text{cyclic},
\end{align}
\end{widetext}
where we have introduced the re-scaled bias parameters
\begin{align}
 \gamma = \frac{b_2}{ b_1} , \qquad
 \gamma' = \frac{b_{s^2}}{ b_1}, \qquad \gamma_* = -\frac{b_{\nabla^2\delta}}{b_1}.   
\end{align}
In writing Eq.~(\ref{bispectrum}) we have neglected stochastic contributions which lead to zero-lag correlations in configuration space, as well as the stochastic noise.

\begin{subsection}{Effective field theory}

An important theoretical ingredient on top of perturbation theory is the effective field theory for large structure formation (EFT) \cite{Baumann:2010tm}. In the standard approach, it arises since the cut-off scale in loop integral regularization requires to be removed by adding counterterms with the appropriate functional form. This prescription effectively models the backreaction of small scales over large scales, with consequences that are of utmost importance for modeling the observed power spectrum and other statistics. However, a different and less discussed approach to introduce the EFT new contributions is to look up directly at the evolution equations and smooth them over some arbitrary scale. By doing so one ends with a theory in which the mass elements act as having internal structure sourcing the Poisson equation with a multipole expansion (see for example \cite{Pietroni:2011iz,Porto:2013qua,Vlah:2015sea}). In the Lagrangian approach to EFT, this scheme induces a correction to the Lagrangian displacement field of the form $\alpha \nabla \delta$ \cite{Vlah:2015sea}.  
Hence, a recipe to add EFT corrections in tree-level statistics is to make the substitution $\delta_g \rightarrow \delta_g + \alpha (k/k_\text{NL})^2 \delta_g$, with the counterterm $\alpha(t)$ considered a free parameter of the theory. However, despite the scales $k_\text{NL}$ and $a\,m(a)$ are not the same and evolve differently with time, the time dependence of $\alpha$ is unknown and hence this EFT counterterm and curvature bias seems to be indistinguishable and degenerate. For this reason, and because at tree-level is more common, we will still use the language of curvature bias when referring to these additions.

Finally, we further mention that EFT counterterms in the tree-level bispectrum are also included to model the non-linear relation between real- and redshift-space coordinate systems along mainly the line-of-sight where the Fingers-Of-God \cite{Ivanov:2021kcd,Philcox:2021kcw}.

\end{subsection}

\end{section}

\begin{section}{Multipole decomposition}\label{subsec:Multipole3PCF}

The three wavevectors entering the bispectrum, $\vk_1$, $\vk_2$ and $\vk_3$, are constrained to form  triangles by statistical homogeneity; and 
because of isotropy, the orientation of these triangles is irrelevant for three-point statistics if redshift-space distortions are not considered, as we do in the rest of the paper.\footnote{For homogeneous but non isotropic fields, one applies these methods to the direction-averaged statistical fields; see e.g.~\cite{Slepian:2015qza,Slepian:2016kfz}, and \cite{Philcox:2021bwo,Philcox:2021hbm} for more general $N$-Point Correlation Functions.} Therefore,
we can characterize these triangles with three numbers, that we choose to be the lengths of two of their sides, $k_1$, $k_2$, and the angle between them $x=\hat{\vk}_1 \cdot \hat{\vk}_2$.
Hence we can write the bispectrum as $B(k_1,k_2,x)$ and expand the internal angle in a Legendre Polynomials $\p_\ell (x)$ basis,
\begin{align} \label{BexpPl}
B(k_1,k_2,x) = \sum_\ell B_\ell(k_1,k_2) \p_\ell (x) 
\end{align}
with
\begin{align}\label{BexpcPl}
 B_\ell(k_1,k_2) = \frac{2 \ell +1}{2} \int_{-1}^1 dx  \p_\ell(x) B(k_1,k_2,x).
\end{align}
The 3PCF is obtained by taking the inverse Fourier transform of Eq.~(\ref{BexpPl}), yielding \cite{Szapudi:2004gg}
\begin{align} \label{zeta}
 \zeta(r_1,r_2,\hat{r}_1\cdot\hat{r}_2) = \sum_{\ell} \zeta_\ell(r_1,r_2) \p_\ell(\hat{r}_1\cdot\hat{r}_2),
\end{align}
where
\begin{align} \label{zetaell}
  \zeta_\ell(r_1,r_2) &= (-1)^\ell  \int \frac{k_1^2k_2^2 dk_1 dk_2}{(2 \pi^2)^2} B_\ell(k_1,k_2)
  \nonumber\\ &\quad
  \qquad \qquad \times j_\ell(k_1 r_1) j_\ell(k_2 r_2).
\end{align}
Notice that the spherical Bessel functions appear when performing the angular piece of the Fourier transform integrals.

The four bias parameters in Eq.~(\ref{halodensity}) combine to give 10 different contributions  to the Legendre multipoles of the 3PCF in Eq.~(\ref{zetaell}). These are $b_1^3$, $b_1^3\gamma$, $b_1^3\gamma'$,
$b_1^3\gamma_*$, $b_1^3\gamma_*\gamma$, $b_1^3\gamma_*\gamma'$, $b_1^3\gamma_*^2$, $b_1^3\gamma_*^2 \gamma$ ,  $b_1^3\gamma_*^2\gamma'$ and $b_1^3\gamma_*^3$. 
% 
% 
% 
% We will separate the bispectrum of Eq.~(\ref{bispectrum}) depending on the bias is applied, finding 10 different contributions
% \begin{align}
% B &= b_1^3 B_{b_1} + b_1^3 \gamma B_{\gamma} + b_1^3 \gamma' B_{\gamma'} + \frac{b_1^3 \gamma_*}{m^2 a^2} B_{\gamma_*} 
%  + \frac{b_1^3 \gamma_* \gamma}{m^2 a^2} B_{\gamma_* \gamma}   + \frac{b_1^3 \gamma_* \gamma'}{m^2 a^2} B_{\gamma_* \gamma'}   \nonumber\\
%   &+ \frac{b_1^3 \gamma_*^2}{m^4 a^4} B_{\gamma_*^2}  
%   + \frac{b_1^3 \gamma_*^2 \gamma}{m^4 a^4} B_{\gamma_*^2\gamma}
%   + \frac{b_1^3 \gamma_*^2 \gamma'}{m^4 a^4} B_{\gamma_*^2\gamma'}
% + \frac{b_1^3 \gamma_*^3}{m^6 a^6} B_{\gamma_*^3}.
% \end{align}
We can reduce this number by noting that we are assuming that only one curvature bias ($\nabla^2 \delta$) 
is sufficient to effectively model the MG biases scale dependence, meaning that $\gamma_*$ is expected to be small. 
Moreover, in ref.~\cite{Aviles:2018saf} it is shown that in MG models
the contributions of curvature bias to the correlation function of tracers is much smaller than those coming from linear and second order local biases. Comparisons to $N$-body simulations \cite{Valogiannis:2019xed,Valogiannis:2019nfz} 
give good fits at large scales, in agreement with the vanishing of higher-order biases. Although this fact is trivial in $\Lambda$CDM, in MG we have introduced the curvature bias in order to effectively account for the scale dependence in the linear local bias; thus, in principle curvature bias can yield considerable contributions at large scales, which are of the same magnitude as the second order local bias. However, we notice that as MG $N$-body simulations became more precise one would be able to measure a non-zero curvature bias. Given this discussion, in the following we will consider terms upto linear order in $\gamma_*$, neglecting quadratic and cubic contributions. In Appendix \ref{app:Bispectrum}, for completeness of the model, we reintroduce the contributions coming from $\gamma_*^2$ and $\gamma_*^3$.

\begin{subsection}{Precyclic bispectrum}

In this section we find the multipoles of the bispectrum performing the cyclic permutations. That is, our goal is to compute
\begin{align}\label{BexpcPlpc}
 B_{pc,\ell}(k_1,k_2) = \frac{2 \ell +1}{2} \int_{-1}^1 dx  B_{pc}(k_1,k_2,x) \p_\ell(x),
\end{align}
with $B_{pc\,,\ell}(k_1,k_2)$ the precyclic ($pc$) multipoles of the bispectrum, which are obtained from Eq.~(\ref{BexpcPl}) without 
considering the cyclic permutations in Eq.~(\ref{bispectrum}). As discussed in the previous section, we are only considering terms up to 
linear order in $\gamma_*$. Hence we split the precyclic bispectrum as
% \begin{align}\label{BpcSplit}
% B_{pc}(k_1,k_2,x) &= b_1^3 B_{pc}^{b_1^3} + b_1^3 \gamma B_{pc}^{b_1^3 \gamma} + b_1^3 \gamma'  B_{pc}^{b_1^3 \gamma'} 
%                    + \frac{b_1^3 \gamma_* }{a^2 m^2} B_{pc}^{b_1^3 \gamma_*} \nonumber\\
%         &\qquad  +  \frac{b_1^3 \gamma_* \gamma}{a^2 m^2} B_{pc}^{b_1^3 \gamma_* \gamma} +  \frac{b_1^3 \gamma_* \gamma'}{a^2 m^2} B_{pc}^{b_1^3 \gamma_* \gamma'} 
% \end{align}
\begin{align}\label{BpcSplit}
\frac{1}{b_1^3} B_{pc}(k_1,k_2,x) &=  B_{pc}^{b_1^3} + \gamma B_{pc}^{ \gamma} + \gamma'  B_{pc}^{ \gamma'}     + \gamma_* B_{pc}^{\gamma_*} \nonumber\\
            &\quad    
        +  \gamma_* \gamma B_{pc}^{\gamma_* \gamma} + \gamma_* \gamma' B_{pc}^{ \gamma_* \gamma'}, 
\end{align}
with each bias combination contribution to the precyclic bispectrum, $B_{pc}^\text{bias type}(k_1,k_2,x)$, given by
\begin{align} 
B_{pc}^{b_1^3} &= 2 F_2(k_1,k_2,x)P_L(k_1)P_L(k_2), \label{BpcSplitcomp1}\\ 
B_{pc}^{ \gamma} &=  P_L(k_1)P_L(k_2), \label{BpcSplitcomp2} \\ 
B_{pc}^{ \gamma'} &= \frac{2}{3} \p_2(x) P_L(k_1)P_L(k_2), \label{BpcSplitcomp3}\\ 
 B_{pc}^{ \gamma_*} &= (k_1^2\!+\!k_2^2\!+\!k_3^2)2 F_2(k_1,k_2,x)P_L(k_1)P_L(k_2),  \label{BpcSplitcomp4}\\ 
B_{pc}^{ \gamma_* \gamma} &= (k_1^2+k_2^2) P_L(k_1)P_L(k_2), \label{BpcSplitcomp5}\\ 
 B_{pc}^{ \gamma_* \gamma'} &= \frac{2}{3} \p_2(x) (k_1^2+k_2^2) P_L(k_1)P_L(k_2). \label{BpcSplitcomp6}
\end{align}
To compute Eq.~(\ref{BexpcPlpc}) for each of the above terms, first, we note that the factors depending only on $k_1$ and $k_2$ in Eqs.~(\ref{BpcSplitcomp1})-(\ref{BpcSplitcomp6}) can be pulled out of the integrals in Eq.~(\ref{BexpcPlpc}). Therefore, for the computation of the multipoles of the precyclic bispectrum one has to obtain
\emph{ 1)} the multipoles of the second order linear bias, which trivially gives $\gamma$ for the monopole and zero for the rest of multipoles,
\emph{ 2)} the tidal contribution, giving $4\gamma'/3$ for the quadrupole and zero otherwise, 
and \emph{ 3)} the multipoles of  the $F_2(k_1,k_2,x)$ and  $k_3^2F_2(k_1,k_2,x)$ kernels. 
The latter becomes equivalent to compute
the multipoles of $xF_2(k_1,k_2,x)$ since the angle cosine $x$ enters through the constriction $k_3^2 = k_1^2 +k_2^2 + 2k_1 k_2 x$. 
For convenience we express the functions $F_2$ and $xF_2$ as
\begin{widetext}
\begin{align}
F_2(k_1,k_2,x) &=  \left(\frac{2}{3} + \frac{3 \mA - \mB}{14} \right) \p_0(x) + \frac{1}{2}\mathcal{G}(k_1,k_2)\p_1(x)  
+ \left(\frac{1}{3} - \frac{1}{7}\mB  \right) \p_2(x), \label{F2PL} \\
 xF_2(k_1,k_2,x) &= \frac{1}{6}\mathcal{G}(k_1,k_2) \p_0(x) + \left[ \frac{4}{5} + \frac{3}{14} \left(\mA - \frac{3}{5}\mB \right)\right] \p_1(x) 
 +\frac{1}{3}\mathcal{G}(k_1,k_2) \p_2(x) \label{xF2PL} + \left(\frac{1}{5} - \frac{3}{35}\mB \right) \p_3(x),
\end{align}
\end{widetext}
where we have introduced the gradient contribution $\mathcal{G}(k_1,k_2) \equiv k_1/k_2 + k_2/k_1$, which arises from transporting large-scale matter bulks along Lagrangian displacement directions. Notice that Eqs.~\eqref{F2PL} and \eqref{xF2PL} are not Legendre multipolar expansions because
$\mA$ and $\mB$ depend on $x$. The exception is 
the $\Lambda$CDM model for which $\mA$ and $\mB$ are only time dependent. Henceforth, in $\Lambda$CDM, the Legendre multipoles that survive in the precyclic bispectrum are $\ell =0,1,2,3$ because of the orthogonal conditions
\begin{align}
\int_{-1}^1 dx \, \p_m(x) \p_n(x) = \frac{2}{2n +1 }\delta_{mn}.  
\end{align}
In MG, instead, all multipoles contribute to the multipolar expansion. However, those
with $\ell>3$ are small since functions $\mA$ and $\mB$ depend weakly on $x$ for fixed wavenumbers $k_1$ and $k_2$. Thus, 
it is natural to split the multipoles of $F_2$ and $xF_2$ as
\begin{align} 
 F_2^{\ell}(k_1,k_2) &= F_{2, \text{LS}}^{\ell}(k_1,k_2) +\Delta F_{2}^{\ell}(k_1,k_2), \label{splitF2} \\
  [x F_2]^{\ell}(k_1,k_2) &= [x F_{2, \text{LS}}]^{\ell}(k_1,k_2) +[\Delta x F_{2}]^{\ell}(k_1,k_2). \label{splitxF2} 
\end{align}
The labels ``LS'' mean that we take the large scale limit of a quantity. 
In models with non-zero mass we have $F_{2, \text{LS}} = F_{2, \text{$\Lambda$CDM}}$, as follows from Eqs.~(\ref{defAk},\ref{DAeveq},\ref{DBeveq}); this is, for example, the case of the $f(R)$ gravity.
Up to the gradient function $\mathcal{G}(k_1,k_2)$, the multipoles $F_{2, \text{LS}}^{\ell}$ are only time dependent functions of order unity, 
while multipoles $\Delta F_2^\ell$ are also $k_1$ and $k_2$ dependent. In Fig.~\ref{fig:F2multipoles} we show
contour plots for the multipoles $\Delta F_{2}^{\ell}(k_1,k_2)$ for the model F4 at redshift $z=0.5$, showing they are small compared to unity, and hence smaller to the
multipoles $F_{2, \text{LS}}^{\ell}$, with the largest contribution coming from $\Delta F_{2}^{\ell=0}(k_1,k_2)< 0.05 F_{2, \text{LS}}^{\ell=0}$. It is important to note that
the overall size of $\Delta F_{2}^{\ell}$ decays quickly with the multipole number $\ell$ which allows us to keep a small number of these terms, in the following we shall consider up to $\Delta F_{2}^{\ell=8}$.

\begin{figure*}
	\begin{center}
	\includegraphics[width=6 in]{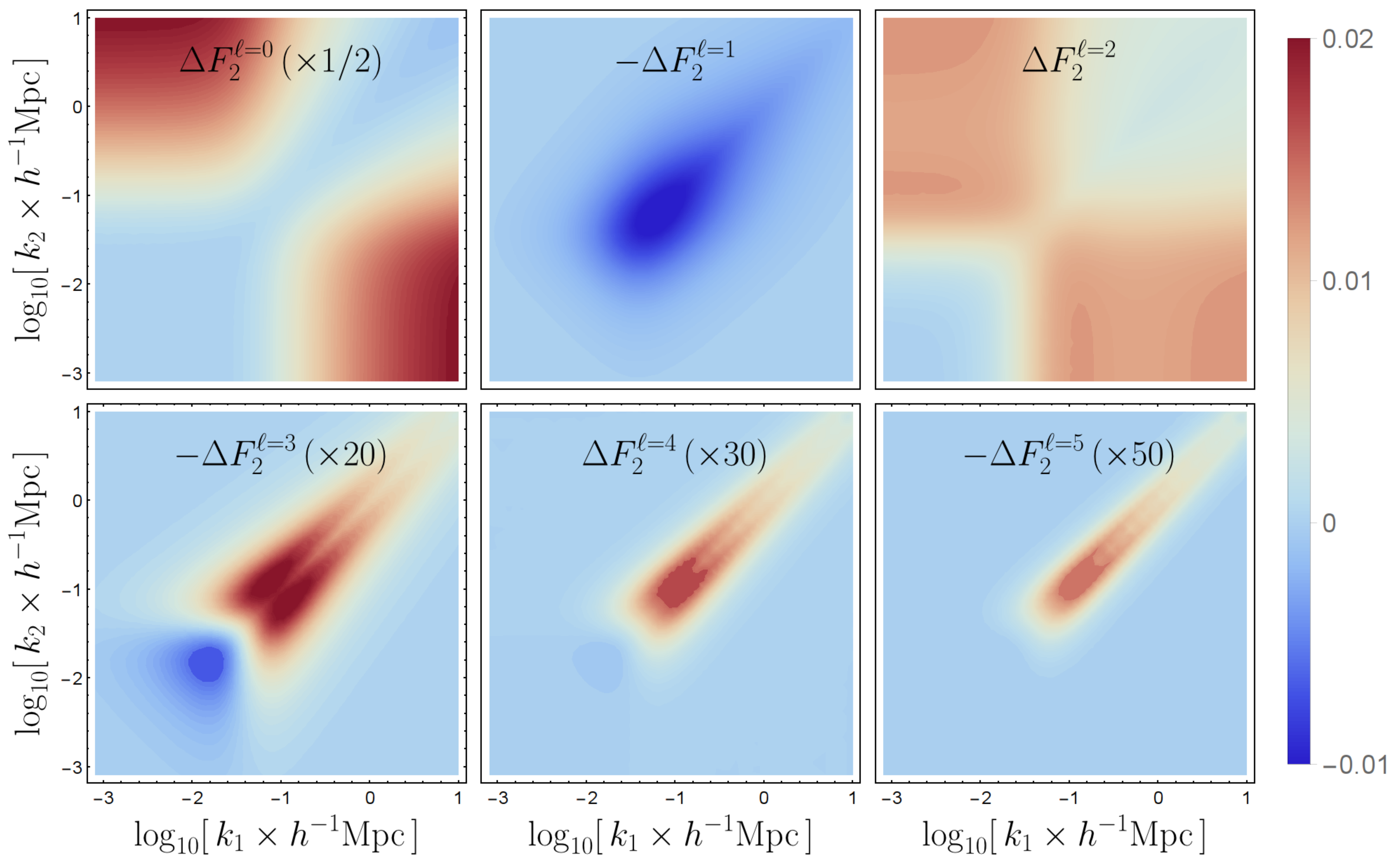}		
	\caption{ $\Delta F_{2}^{\ell}(k_1,k_2)$ for F4 model at redshift $z=0.5$. All functions
	we alternate sign according to the multipole number $\ell$ because this how they appear in 
	the pre-cyclic 3PCF of Eq.~\eqref{zetaellpc}
	\label{fig:F2multipoles}}
	\end{center}
\end{figure*}

In models with non-vanishing mass, such as in $f(R)$ theories,
the computation of multipoles $F_{2,\text{LS}}^\ell$  can be performed by letting $k_1$ and $k_2$ go to zero in $F_{2}^\ell(k_1,k_2)$,
as stated above. These are given by
\begin{align}
  F_{2, \text{LS}}^{\ell=0} &= \frac{14+3\mA^\text{LS}}{21}, \qquad 
  F_{2, \text{LS}}^{\ell=1} = \frac{1}{2}\mathcal{G}(k_1,k_2), \\
  F_{2, \text{LS}}^{\ell=2} &= \frac{7-3 \mA^\text{LS}}{21}, \qquad
  F_{2, \text{LS}}^{\ell>2} = 0.
\end{align}
The standard, well known values computed with EdS kernels (see e.g. \cite{Slepian:2016weg}),
$ F_{2, \text{EdS}}^{\ell=0}=17/21$ and $ F_{2, \text{EdS}}^{\ell=2}=4/21$, are recovered by setting $\mA^{ \text{LS}}=1$.
Equivalently, the multipoles of $x\, F_{2, \text{LS}}$ are
\begin{align}
[x\, F_{2, \text{LS}}]^{\ell=0} &= \frac{1}{6} \mathcal{G}(k_1,k_2),\qquad 
[x\, F_{2, \text{LS}}]^{\ell=1} = \frac{28 + 3 \mA^{ \text{LS}}}{35}, \\
[x\, F_{2, \text{LS}}]^{\ell=2} &= \frac{1}{3} \mathcal{G}(k_1,k_2), \qquad
[x\, F_{2, \text{LS}}]^{\ell=3} =  \frac{7- 3 \mA^{ \text{LS}}}{35}, \\
[x\, F_{2, \text{LS}}]^{\ell>3} &= 0.
\end{align}
Meanwhile, the multipoles of $\Delta F_2(k_1,k_2,x)$ are
% \begin{align}
% [\Delta F_2]^{\ell}(k_1,k_2) &=  \frac{1}{14}[3 \mA - \mB(1 + 2 \p_2)]^{\ell}  -  \frac{\mA^{ \text{LS}}}{7} ( \delta_{\ell0} -\delta_{\ell 2} ), \\
%  [\Delta (x F_2)]^\ell(k_1,k_2) &= \frac{3}{70}[ 5 \mA \p_1 - \mB (3\p_1+2\p_3)]^{\ell}   
%  -\frac{3 \mA^{ \text{LS}}}{35}(\delta_{\ell 1}-\delta_{\ell3} ). 
% \end{align}
\begin{align}
[\Delta F_2]^{\ell}(k_1,k_2) &=  \frac{3}{14}[ \mA - \mB x^2]^{\ell}  -  \frac{\mA^{ \text{LS}}}{7} ( \delta_{\ell0} -\delta_{\ell 2} ), \\
 [\Delta (x F_2)]^\ell(k_1,k_2) &= \frac{3}{14}[  \mA x - \mB x^3]^{\ell}   
 -\frac{3 \mA^{ \text{LS}}}{35}(\delta_{\ell 1}-\delta_{\ell3} ), 
\end{align}
which should be computed numerically.

For the DGP models, the scale dependencies come from the Vainshtein screening which affects only the monopole and quadrupole of the bispectrum in
virtue of Eq.~(\ref{sourceSDGP}). 
For notational consistency, and only in the DGP model case, we will refer as $\mA_\text{LS}$ to the function $\mA(t)$ obtained without considering the 
screening source $S_2$ in Eq.~(\ref{DAeveq}), and $\Delta F_2 = F_2|_{S_2\neq 0 } - F_2|_{S_2= 0 }$. In this way we can still use the splitting
$F_2 = F_{2,\text{LS}} + \Delta F_2$, but in DGP the first term refers to the kernel  in the absence of screenings, 
while the second term are the screening contributions.
The screening source for DGP [Eq.~(\ref{sourceSDGP})]
can be written as $S_2(x,t) = \frac{2}{3}Z(t) (\p_0(x)- \p_2(x))$ and therefore 
\begin{align}
\Delta F_2^\ell = f(t)( -\delta_{l0}+\delta_{l2}) 
\end{align}
with the time dependent function $f(t)$ obtained by solving
\begin{align} \label{DGPf}
 f(t) = \frac{1}{3} D_+^{-2}(t) \big(\T - A(t) \big)^{-1} \big[ Z(t)D_+^2(t) \big]. 
\end{align}
In such a way we can use all of the above formulae also in the case of DGP.
% The function $f(t)$ depends on the specific model and it is obtained by solving the Green problem $f(t)=(\T-A(t))^{-1}g(t)$;  
% in particular for N5 and N1 these are $f(t_0)\simeq 0.0017$ and $f(t_0) \simeq 0.0044$, respectively, for a background cosmology $\Omega_{m0}= 0.281$, $h=0.697$. 
In particular, for N5 and N1 function $f$ takes values $f(t_0)\simeq 0.002$ and $f(t_0) \simeq 0.005$, respectively, for a background cosmology $\Omega_{m0} \simeq 0.3$ and $h=0.7$. Hence, similar plots than those in 
Fig.~\ref{fig:F2multipoles} but for DGP, will show non-zero values only for multipoles $\ell=0$ and $\ell=2$, and these are
simply constants in the $k_1$-$k_2$ space.

With these partial results, we find the following bispectrum multipole components
\begin{align}\label{Bpclincomplete}
B^{b_1^3}_{pc,\ell}(k_1,k_2) &= 2 P_L(k_1) P_L(k_2) \left( F_{2,\text{LS}}^{\ell} + [\Delta F_2]^{\ell} \right), \nonumber\\
B^{\gamma}_{pc,\ell}(k_1,k_2) &= P_L(k_1) P_L(k_2) \delta_{\ell 0}, \nonumber\\
B^{\gamma'}_{pc,\ell}(k_1,k_2) &= \frac{4}{3} P_L(k_1) P_L(k_2) \delta_{\ell 2}.
\end{align}
The multipoles for the remaining three components considered in this work, 
$B^{\gamma_*}$, $B^{\gamma_*\gamma}$ and $B^{\gamma_*\gamma'}$, 
whose expressions are cumbersome, are given in Appendix \ref{app:Bispectrum}.

Before continuing with the computation of the 3PCF, it is worth discussing the differences in behavior in different cosmological models. For the EdS case, the function $\mA = \mA_\text{LS}$ is equal to unity while the kernel $\Delta F_2$ is zero. 
Hence $F_2^{\ell=0}$ and $F_2^{\ell=2}$, which correspond to the evolution of spherical collapse dynamics 
and the effect of tidal gravitational fields, respectively, remain constant. 
On the other hand, for $\Lambda$CDM $\Delta F_2$ is still vanishing, and one has $\mA_\text{LS}>\mA_\text{EdS}$, which makes
the monopole larger than its EdS value, but also a smaller quadrupole. 
%Which is showing that spherical collapse is most efficient in MG theories, as expected from earlier works on the halo mass function, while tidal gravitational contributions are attenuated. This last point should not be a surprise 
This implies that, even for $\Lambda$CDM, the different multipoles of the bispectrum, and 
consequently of the 3PCF, grow at different rates, contrary to the EdS case in which all multipoles grow simply as 
$D_+^4(t)$.\footnote{This is the common approach in PT, which uses EdS kernels, 
but the precise linear growing functions in $\Lambda$CDM.} 
For $f(R)$ theories, this effect is enhanced since we have to sum both contributions 
$\Delta F_2^{\ell=0}$ and $\Delta F_2^{\ell=2}$, which are positive (as can be read 
from Fig.~\ref{fig:F2multipoles}), but notice the monopole term is about two times larger than the quadrupole. This is consistent with earlier finding that the spherical collapse and the formation of halos is much more efficient in $f(R)$ than in  $\Lambda$CDM \cite{Li:2011qda,Kopp:2013lea}. 
Furthermore, we have a contribution for $\Delta F_2^{\ell=1}$, not present in $\Lambda$CDM or DGP. 
This dipole comes from linear displacements of fluid positions, and hence is affected in massive theories, 
for which the standard Lagrangian displacement-overdensity relation receives additional scale dependencies 
through the linear growth function $D_+(k,t)$. The relation is 
\begin{equation}
    \Psi_i(\vk,t)=i\frac{k_i}{k^2}D_+(k,t)\delta(\vk,t),
\end{equation}
with $\Psi_i$ the Lagrangian displacement field to first order in PT. 

In DGP the situation is rather
different, here  $\mB^\text{DGP} = \mA^\text{DGP}_\text{LS}>\mA_\text{$\Lambda$CDM}$, 
and hence $F^{\ell=0}_\text{2,DGP}>F^{\ell=0}_\text{2,$\Lambda$CDM}$ and $F^{\ell=2}_\text{2,DGP}<F^{\ell=2}_\text{2,$\Lambda$CDM}$. 
But, as we have seen above, the contributions of $\Delta F_2$ are negative for
the monopole and positive for the quadrupole,  driving the whole $F_2$ kernel to that of $\Lambda$CDM. This is a consequence of
the new scale dependencies in DGP enter only 
through nonlinearities of the Klein-Gordon equation, and as a result, the $\Delta F_2^{\ell}$ are pure screening contributions in DGP.

\end{subsection}

\begin{subsection}{Precyclic 3PCF}

With the $B_{\ell,pc}$ pieces at hand, we can Fourier transform them to obtain the precyclic 3PCF multipoles $\zeta_{pc\,,\ell}$. That is, we aim to compute
%\begin{widetext}
\begin{align} \label{zetaellpc}
  \zeta_{pc\,,\ell}(r_1,r_2) &\equiv   (-1)^\ell \int \frac{k_1^2k_2^2 dk_1 dk_2}{(2 \pi^2)^2} B_{pc\, ,\ell}(k_1,k_2) \nonumber\\ &\quad \times j_\ell(k_1 r_1) j_\ell(k_2 r_2).
\end{align}
%\end{widetext}
If the precyclic $B_\ell$ are separable in $k_1$ and $k_2$ dependent factors, the integrals of the precyclic 3PCF multipoles in Eq.~(\ref{zetaellpc}) 
reduce to the multiplication of two 1-dimensional integrals. This is the case of
$\Lambda$CDM in real space \cite{Slepian:2014dda,Slepian:2016weg}. For MG, however, the only pieces of $B_\ell$ that are separable are those with $\gamma$ and $\gamma'$ biasing factors. However, the splitting of $F_2$ and 
$xF_2$ in Eqs.~(\ref{splitF2}) and (\ref{splitxF2}) 
allow for the pieces containing multipoles of $ F_{2,  \text{LS}}$ and $x F_{2,\text{LS}}$ to also be separable, since these are scale independent.

To obtain the multipoles of the 3PCF using Eq.~(\ref{zetaellpc}), it is convenient to introduce different definitions. To begin, we use the new notation
\begin{align}
\xi^{[n,m]}(r) &= \int \frac{k^2 dk}{2 \pi^2}k^{m} P_L(k) j_n(k r), \\  
\xi^{[n]}(r) &= \xi^{[n,0]}(r).
\end{align}
Note that $\xi^{[0]}(r)=\xi_L(r)$ is simply the linear correlation function. We further introduce the functions
\begin{align}
 X_0(r_1,r_2) &= \xi^{[0]}(r_1)\xi^{[0]}(r_2),  \\
 X_1(r_1,r_2) &= \xi^{[1,1]}(r_1)\xi^{[1,-1]}(r_2) \nonumber\\
              &\quad + \xi^{[1,-1]}(r_1)\xi^{[1,1]}(r_2),  \\
 X_2(r_1,r_2) &= \xi^{[2]}(r_1)\xi^{[2]}(r_2),  %\\
%  X_3(r_1,r_2) &=  \xi^{[3,1]}(r_1)\xi^{[3,-1]}(r_2) + \xi^{[1,-1]}(r_1)\xi^{[3,1]}(r_2),  
\end{align}
that show up after the multipoles $F_{2,\text{LS}}^{\ell=0,1,2}$ are integrated, as can be checked straightforwardly.  We finally define
% enter into the local and tidal biasing terms and 
\begin{align}
& Y_0(r_1,r_2) =  \xi^{[0,2]}(r_1) \xi^{[0]}(r_2) +  \xi^{[0]}(r_1) \xi^{[0,2]}(r_2),  \\
&Y_1(r_1,r_2) =  \xi^{[1,1]}(r_1) \xi^{[1,1]}(r_2) + \frac{35}{98 + 3  \mA^{ \text{LS}}}\nonumber\\ 
&\quad   \times \Big[ \xi^{[1,3]}(r_1) \xi^{[1,-1]}(r_2) +\xi^{[1,-1]}(r_1) \xi^{[1,3]}(r_2) \Big], \\
& Y_2(r_1,r_2) = \xi^{[2,2]}(r_1) \xi^{[2]}(r_2) +  \xi^{[2]}(r_1) \xi^{[2,2]}(r_2), \\
& Y_3(r_1,r_2) = \xi^{[3,1]}(r_1) \xi^{[3,1]}(r_2),
\end{align}
that will enter when integrating the multipoles $[x F_{2,\text{LS}}]^{\ell=0,1,2,3}$. These $Y$ functions correspond to 
curvature biasing terms.

\begin{figure*}
	\begin{center}
	\includegraphics[width=7 in]{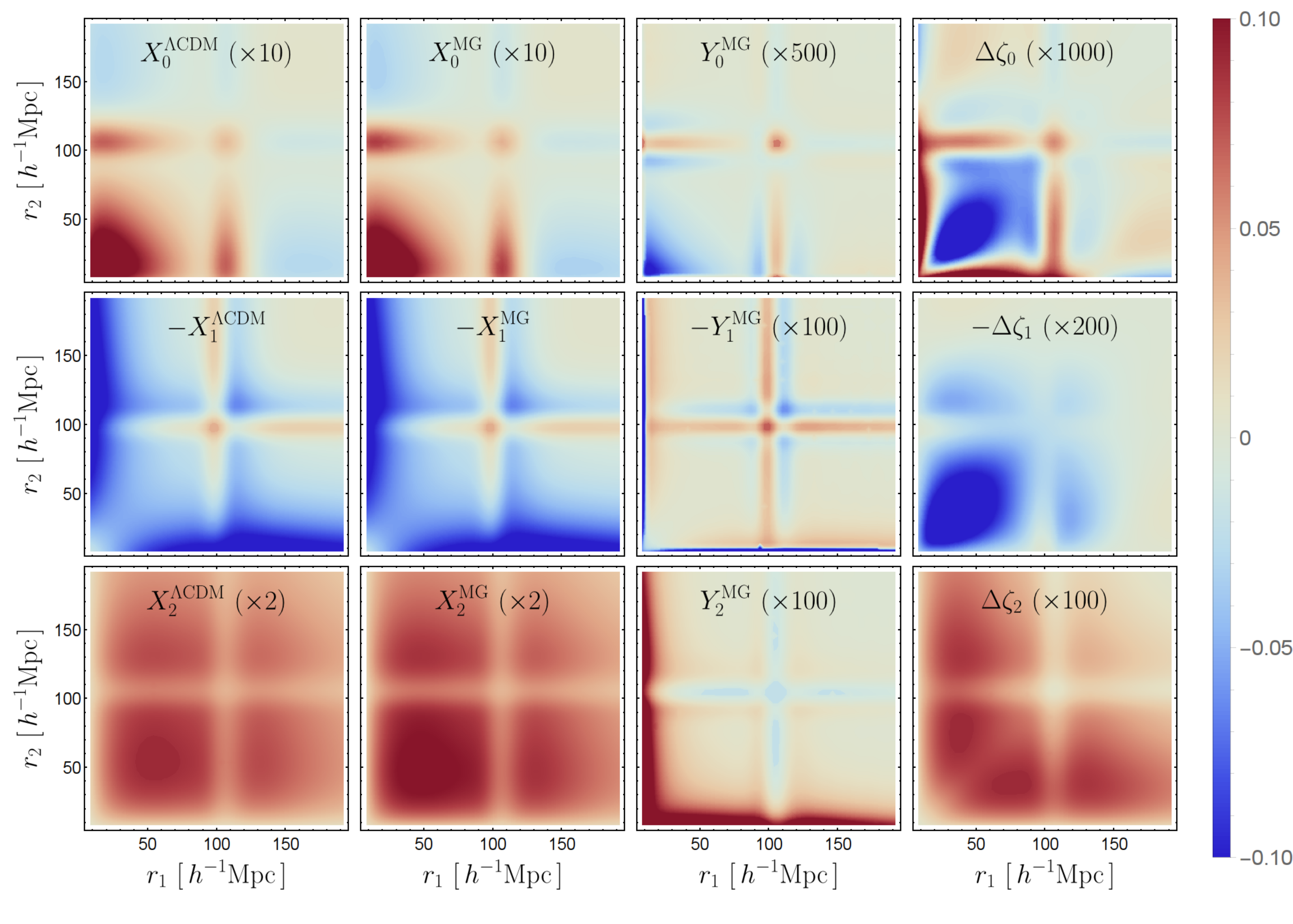}		
	\caption{Pre-cyclic contributions to the 3PCF in Eq.~(\ref{pczetaell}) for models $\Lambda$CDM and F4 at redshift $z=0.5$. All functions
	are multiplied by $r_1^2 r_2^2/(10 \,h^{-1}\text{Mpc})^4$. The $X$ and $Y$ functions alternate signs because this how they appear in 
	the pre-cyclic 3PCF due to the prefactor $(-1)^\ell$ in Eq.~(\ref{pczetaell})
	\label{fig:prec3pcflm}}
	\end{center}
\end{figure*}

\begin{figure*}
	\begin{center}
	\includegraphics[width=5 in]{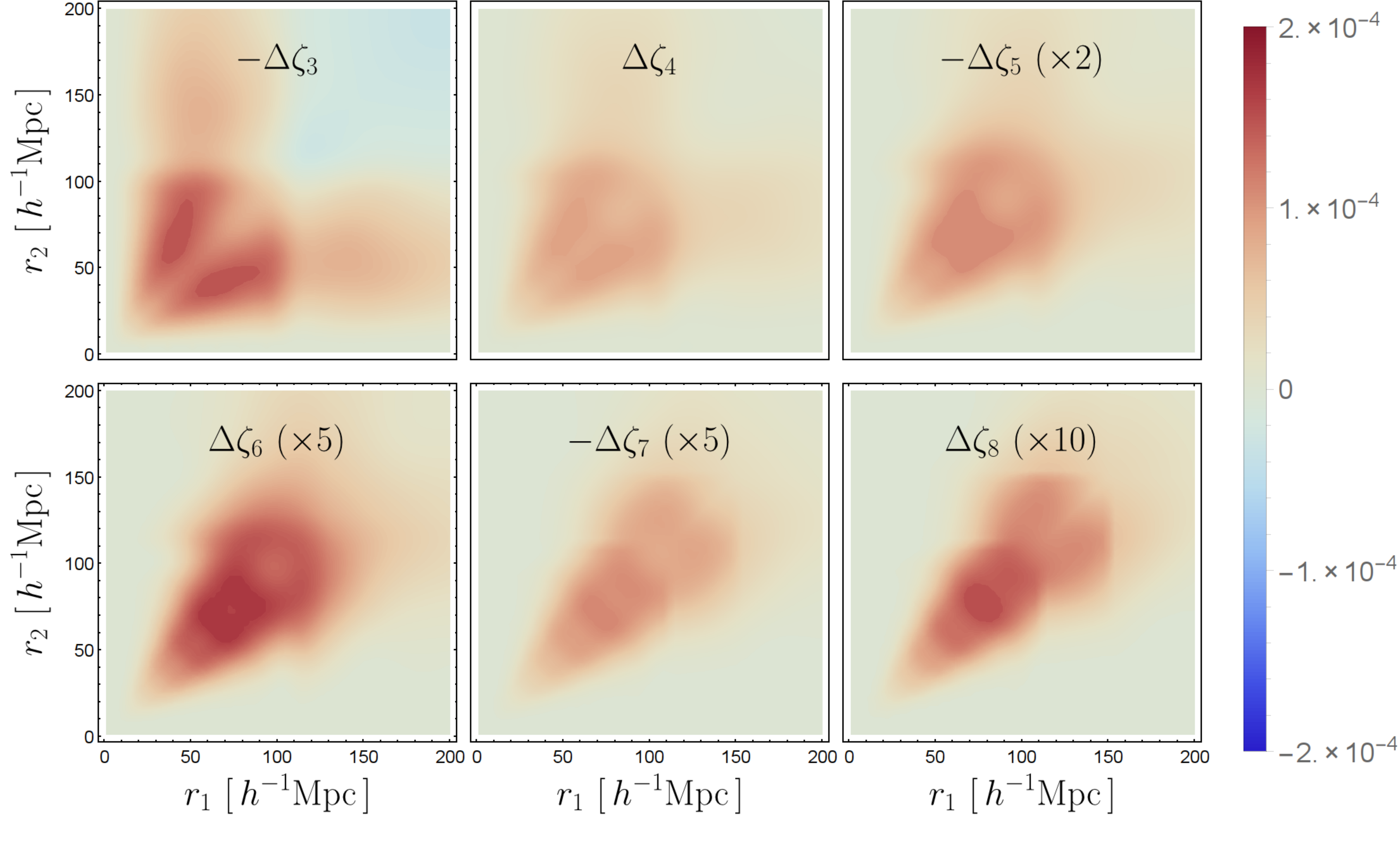}		
	\caption{Pure MG contributions $\Delta \zeta_\ell$ to the precyclic 3PCF given by  Eq.~\eqref{Deltazetaell}. We use the F4 model at redshift $z=0.5$. All functions
	are weighted by $r_1^2 r_2^2/(10 \text{Mpc}/ h)^4$.
	\label{fig:prec3pcfhm}}
	\end{center}
\end{figure*}

Inserting the bispectrum biasing components  
[Eqs.~(\ref{Bpclincomplete}) and Eqs.~(\ref{Bgsell0}-\ref{Bgsell3}, \ref{Bgsgell}, \ref{Bgsgpell}) of Appendix \ref{app:Bispectrum}] into Eq.~(\ref{zetaellpc}),
and using the definitions of $X_\ell$ and $Y_\ell$ functions, we obtain the precyclic 3PCF multipoles, namely
% \begin{align}
% \zeta_0(r_1,r_2) &=  b^3_1 \int \frac{k_1^2k_2^2 dk_1 dk_2}{(2 \pi^2)^2} 2 F_2^{\ell=0}(k_1,k_2)P(k_1)P(k_2)j_0(k_1 r_1) j_0(k_2 r_2) + 
% b^3_1 \gamma \xi^{[0]}(r_1)\xi^{[0]}(r_2) \label{pc3pcfl0}\\
% \zeta_2(r_1,r_2) &=  b^3_1 \int \frac{k_1^2k_2^2 dk_1 dk_2}{(2 \pi^2)^2} 2 F_2^{\ell=2}(k_1,k_2)P(k_1)P(k_2)j_2(k_1 r_1) j_2(k_2 r_2) + 
% \frac{2}{3} b^3_1 \gamma' \xi^{[2]}(r_1)\xi^{[2]}(r_2) \label{pc3pcfl2}\\
% \zeta_\ell(r_1,r_2) &=  (-1)^\ell b^3_1 \int \frac{k_1^2k_2^2 dk_1 dk_2}{(2 \pi^2)^2} 2 F_2^{\ell}(k_1,k_2)P(k_1)P(k_2)j_\ell(k_1 r_1) j_\ell(k_2 r_2) 
% \qquad (\ell \neq 0,2) \label{pc3pcfl}\\
% \end{align}
% \begin{align}
% \zeta_0(r_1,r_2) &=  b^3_1 \big( 2 F_{2,  \text{LS}}^{\ell=0} + \gamma \big)X_0(r_1,r_2) + 
%  b^3_1 \Delta \zeta_0(r_1,r_2), \label{pc3pcfl0-0}\\
% \zeta_1(r_1,r_2) &=  - b^3_1  X_1(r_1,r_2) +  b^3_1 \Delta \zeta_1(r_1,r_2) \label{pc3pcfl1-0}\\
% \zeta_2(r_1,r_2) &=  b^3_1 \left( 2 F_{2,  \text{LS}}^{\ell=2} + \frac{2}{3}\gamma'  \right) X_2(r_1,r_2) + 
%  b^3_1 \Delta \zeta_2(r_1,r_2), \label{pc3pcfl2-0}\\
% \zeta_\ell(r_1,r_2) &=  b^3_1 \Delta \zeta_\ell(r_1,r_2)  \qquad (\ell >2), \label{pc3pcfl-0}
% \end{align}
% with the barred 3pcf 
\begin{widetext}
\begin{align}\label{pczetaell}
\zeta_{pc,\ell} (r_1,r_2) &= b_1^3 \left( x_\ell + \gamma \delta_{\ell0}+  \frac{4}{3}\gamma' \delta_{\ell 2}\right) X_\ell(r_1,r_2) + b^3_1 \Delta \zeta_\ell(r_1,r_2)  + \frac{ b_1^3 \gamma_*}{a^2 m^2} \left( y_\ell + \gamma \delta_{\ell0} +\frac{4}{3}\gamma' \delta_{\ell 2}\right) Y_\ell(r_1,r_2),
\end{align}
\end{widetext}
with $x_\ell(t)$ and $y_\ell(t)$ functions depending weakly on time and given by
\begin{align}
 x_0 = \frac{28+ 6 \mA^{ \text{LS}}}{21},&\quad x_1=-1,\quad x_2=   \frac{14-6\mA^{ \text{LS}}}{21},\\
 y_0 = \frac{70+12\mA^{ \text{LS}}}{21} ,&\quad  y_1= -\frac{392 + 12 \mA^{ \text{LS}}}{35}, \\ 
  y_2= \frac{56 - 12 \mA^{ \text{LS}}}{21},&\quad y_3 = -\frac{28- 12 \mA^{ \text{LS}}}{35}, 
\end{align}
while $x_{\ell>2}=0$ and $y_{\ell>3}=0$. Furthermore, the configuration space counterpart of the $\Delta F_2$ contribution results in
\begin{align}\label{Deltazetaell}
\Delta \zeta_\ell(r_1,r_2) &\equiv 2  (-1)^\ell  \int \frac{k_1^2 \,dk_1}{2 \pi^2} \frac{k_2^2\,dk_2}{2 \pi^2} \Delta F_2^{\ell}(k_1,k_2) \nonumber\\
&\quad \times  P_L(k_1)P_L(k_2)j_{\ell}(k_1 r_1) j_{\ell}(k_2 r_2). 
\end{align}
%{\bf ZS: in 3.59 above is the factor of 2 before $\Delta$ intended?} \alej{Yes, it comes from $B= 2 F_2 P_L^2 + cyclic$}

In Fig.~\ref{fig:prec3pcflm} we show 2-dimensional plots for the $X_\ell$ and  $Y_\ell$ functions for multipoles $\ell=0,1,2$ and for $\Lambda$CDM and the HS F4 model. The contributions from $X$ functions, which are related to linear and quadratic local and tidal biases are dominant, hence our choice of neglect beyond linear $\gamma^*$ is well justified. Notice we do not show $Y_\ell$ functions for $\Lambda$CDM, since the bias expansion with $b_1$, $b_2$ and $b_{s^2}$ is complete up to the second order, which is not the case in $f(R)$.  In the last column of this same figure we show the contributions $\Delta \zeta_\ell$ given by Eq.~\eqref{Deltazetaell}, which are purely MG effects.  Notice that despite the scaling factors, the discrepancy with GR can be as large as the $2\%$ of the $X_\ell$ functions at the intermediate scales $r_1 \sim r_2 \sim 50 \,h^{-1}\text{Mpc}$. Moreover, for the precyclic $\Lambda$CDM 3PCF the description ends at $\ell = 2$. In contrast, for MG the terms due to curvature bias introduce an $\ell=3$ contribution, while the $\Delta F_2$ piece yields to infinite multipoles for $\Delta \zeta_\ell$. A few of the latter are plotted in Fig.~\ref{fig:prec3pcfhm} where, fortunately but somewhat expected, their overall amplitudes decay quickly with the multipole $\ell$.

As a remark, the larger effects of MG in Eq.~(\ref{pczetaell}) come from the linear power spectrum as well as from the $\Delta \zeta_\ell$ contributions.  
There are additional corrections coming from a similar expression to Eq.~(\ref{Deltazetaell}) with $\Delta F_2^{\ell}$ replaced by $[\Delta x F_2]^{\ell}$. However, these terms are even smaller than the $Y_\ell$ functions ---which are similar in size to the terms $\Delta \zeta_\ell$--- and
given that they are further multiplied by $\gamma_*$, which is assumed to be small, we neglect them.

In DGP, as shown in the previous section, only the monopole and quadrupole of $\Delta F_2$ survive and these are scale independent. Hence,
the 2-dimensional integral of Eq.~(\ref{Deltazetaell}) reduces to the product of two 1-dimensional integrals. For the monopole one obtains 
$\Delta \zeta_{\ell=0} \propto X_0$, for the quadrupole $\Delta \zeta_{\ell=2} \propto X_2$, and zero otherwise. Hence we can simplify the 
expression (\ref{pczetaell}) by absorbing these contributions into functions $x_{0,2}(t)$ with the substitutions $x_0 \rightarrow x_0 + 2f(t)$,  $x_2 \rightarrow x_2 - 2f(t)$, with $f$ given by Eq.~(\ref{DGPf}), and setting $\Delta \zeta_{\ell}=0$,
showing that the Vainshtein screening in DGP yields a signal close to the percent level in the 3PCF monopole and quadrupole, while no effects into other multipoles. This situation is similar in other theories with a Vainshtein mechanism. Indeed, the structure of the nonlinear derivative terms in Eq.~(\ref{sourceSDGP}) is shared by other models, such as the cubic Galileons \cite{Nicolis:2008in} or certain sectors of Hordenski (see for example \cite{Koyama:2013paa}) , and therefore
such theories will show up the same qualitative behaviour. 

\end{subsection}

\begin{subsection}{The 3PCF multipoles}

\begin{figure*}
	\begin{center}
	\includegraphics[width=7 in]{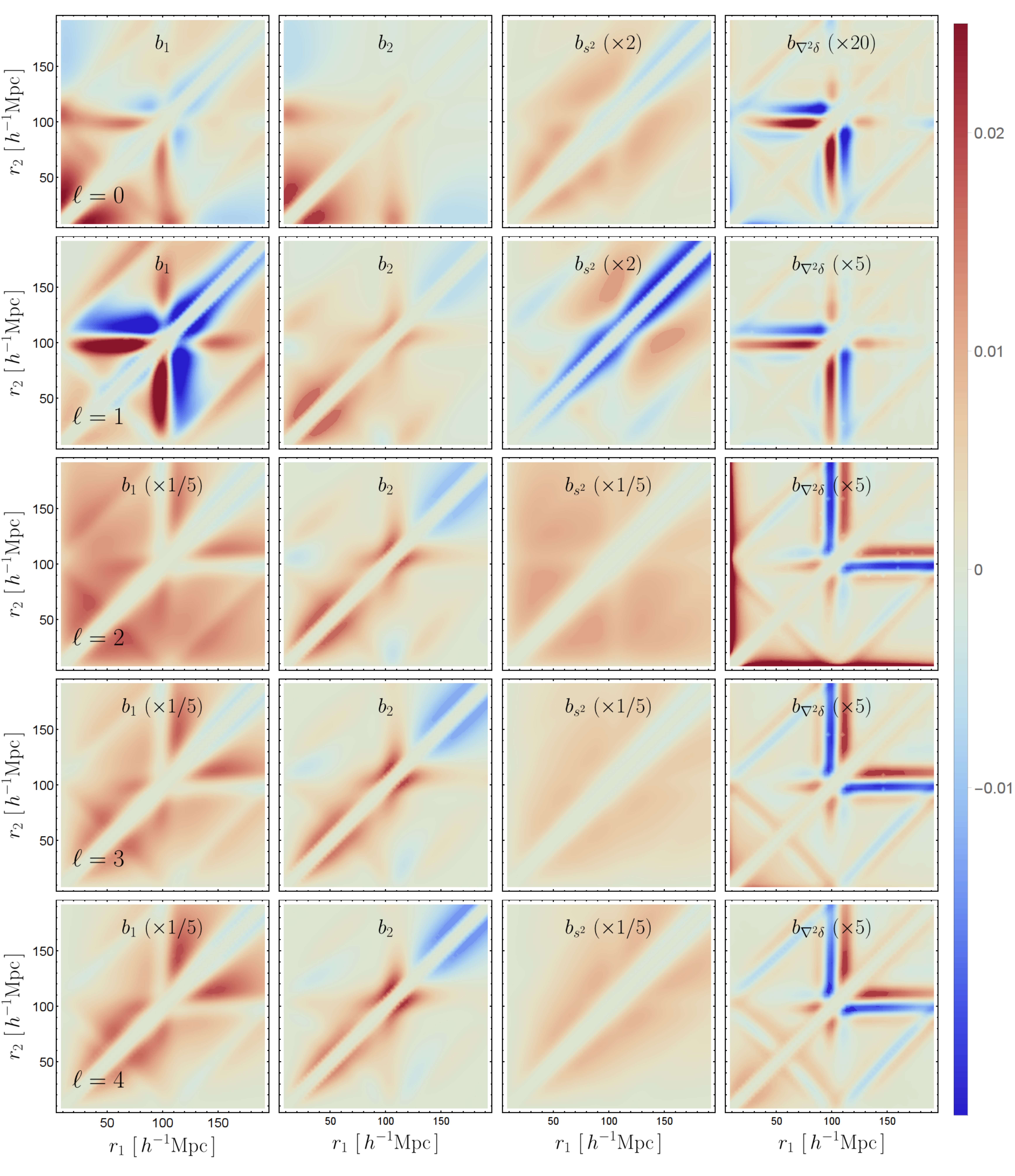}		
	\caption{Contributions to the 3PCF in Eq.~(\ref{pstcyc3PCF}) for F4 model at redshift $z=0.5$. We show the contributions due to biases $b_1$, $b_2=b_1 \gamma$, $b_s = b_1 \gamma'$, 
	and $b_{\nabla^2\delta} = b_1 \gamma_*$ from left to right and for multipoles $\ell=0,1,2,3,4$ from top to bottom. 
	All functions
	are multiplied by $r_1^2 r_2^2/(10 \,h^{-1} \text{Mpc})^4$ and smoothed by a function $\exp[-(12\,h^{-1}\text{Mpc}  /(r_1-r_2))^2]$.
	\label{fig:postc3pcflm}}
	\end{center}
\end{figure*}

The total 3PCF is obtained by cyclic summing the pre-cyclic 3PCF piece previously obtained, namely
\begin{align}\label{postc3pcf}
 &\zeta(r_1,r_2,\hat{r}_1\cdot\hat{r}_2) = \sum_L  \big[\zeta_{pc,L}(r_1,r_2) \p_L(x_{12}) \nonumber\\
 &\quad + \zeta_{pc,L}(r_2,r_3) \p_L(x_{23}) + \zeta_{pc,L}(r_3,r_1) \p_L(x_{31}) \big],
\end{align}
where $r_3$, $x_{23}$ and $x_{31}$ can be written as functions of $r_1$, $r_2$ and $x_{12}$. The labels $pc$ indicate that we refer to the pre-cyclic functions of the previous section. 
The Legendre multipoles of the post-cyclic 3PCF are
\begin{align}\label{postc3pcfmult}
 \zeta_\ell(r_1,r_2) = \frac{2 \ell + 1}{2} \int_{-1}^1 d x_{12} \,\zeta(r_1,r_2,x_{12}) \p_\ell(x_{12}).
\end{align} 
Moreover, in order to track the different contributions, we define the following projections of pre-cyclic multipoles $L$ onto multipole $\ell$
\begin{align}
 \mI^{(\ell,L)}(r_1,r_2) &\equiv \frac{2 \ell + 1}{2} \int_{-1}^{1} dx_{12} \Big[ X_L(r_2,r_3)\p_L(x_{23})  \nonumber\\
 &\quad + X_L(r_3,r_1)\p_L(x_{31}) \Big] \p_\ell(x_{12}), \\
 \Delta \mI^{(\ell,L)}(r_1,r_2) &\equiv \frac{2 \ell + 1}{2} \int_{-1}^{1} dx_{12} \Big[ \Delta \zeta_L(r_2,r_3)\p_L(x_{23}) \nonumber\\
 &\quad
 + \Delta \zeta_L (r_3,r_1)\p_L(x_{31})   \Big] \p_\ell(x_{12}), \\
J^{(\ell,L)}(r_1,r_2) &= \frac{2 \ell + 1 }{2}\int_{-1}^1 dx_{12} \Big[ Y_L(r_2,r_3) \p_L(x_{23}) \nonumber\\
 &\quad+ Y_L(r_3,r_1) \p_L(x_{31})  \Big] \p_\ell(x_{12}). 
\end{align}
From these $\mI$ and $J$ functions we obtain the postcyclic 3PCF 
\begin{widetext}
\begin{align} \label{pstcyc3PCF}
\zeta_\ell(r_1,r_2) &= \zeta_{pc,\ell}(r_1,r_2) + b_1^3 \sum_{L=0}^2 \left( x_L + \gamma\delta_{\ell 0} + \frac{4}{3} \gamma'\delta_{\ell 2}\right) \mI^{(\ell,L)}(r_1,r_2) 
 + b_1^3 \sum_{L=0}^\infty  \Delta \mI^{(\ell,L)}(r_1,r_2) \nonumber\\
 &\quad +  b_1^3 \gamma_*\sum_{L=0}^3 \left( y_L + \gamma\delta_{\ell 0} + \frac{4}{3} \gamma'\delta_{\ell 2}\right) J^{(\ell,L)}(r_1,r_2).
\end{align}
\end{widetext}
This equation describes the complete tree-level 3PCF of biased in the Szapudi-basis. 

In Fig.~\ref{fig:postc3pcflm}, we show the contributions for multipoles $\ell=0,\dots,4$ in the HS F4 model at redshift $z=0.5$. To cover properly the range of the multipoles over the chosen domain we multiply them by $r_1^2 r_2^2 \,(10 \,h^{-1} \text{Mpc})^{-4}$, where the numerical factor is chose to deal with dimensionless quantities.  We show the contours to the different 
biasing terms  to the above equation with the exception of $b_1 \gamma_*\gamma $ and $b_1 \gamma_*\gamma$. 
To obtain the total post-cyclic 3PCF for biased tracers one must sum each column weighted by the corresponding bias parameters using eq.~\eqref{pstcyc3PCF}. For visualization purposes, we have smoothed the contours with an exponential function $\exp[-(12\, h^{-1} \text{Mpc} /(r_1-r_2))^2]$. This is because the dominant contribution comes from the diagonal $r_1 \simeq r_2$, corresponding to isosceles triangles and hence the closing side of the triangle ($r_3$) can be arbitrarily small and get out of the reach of PT, and as such, not well modeled by our method.  We notice that contrary to the precyclic results, is not clear that the amplitude of the multipoles start to decrease beyond some $\ell$, actually, it seems to be similar for $\ell \geq 2$. In the literature there is not a formal proof that this happens even in the case of $\Lambda$CDM, and if convergence is not attained that would mean that we cannot reconstruct the whole 3PCF from its multipoles. However, this is not a significant obstacle to use them, since their advantage rely in that the computational complexity to get these statistics from the data is reduced drastically and the estimators to do so search directly for the multipoles and not for the whole 3PCF \cite{Slepian:2015qza,Philcox:2021bwo}.

\end{subsection}

\end{section}

\end{section}

\begin{subsection}{Application to halo catalogues}

\begin{figure*}
	\begin{center}
	\includegraphics[width=5 in]{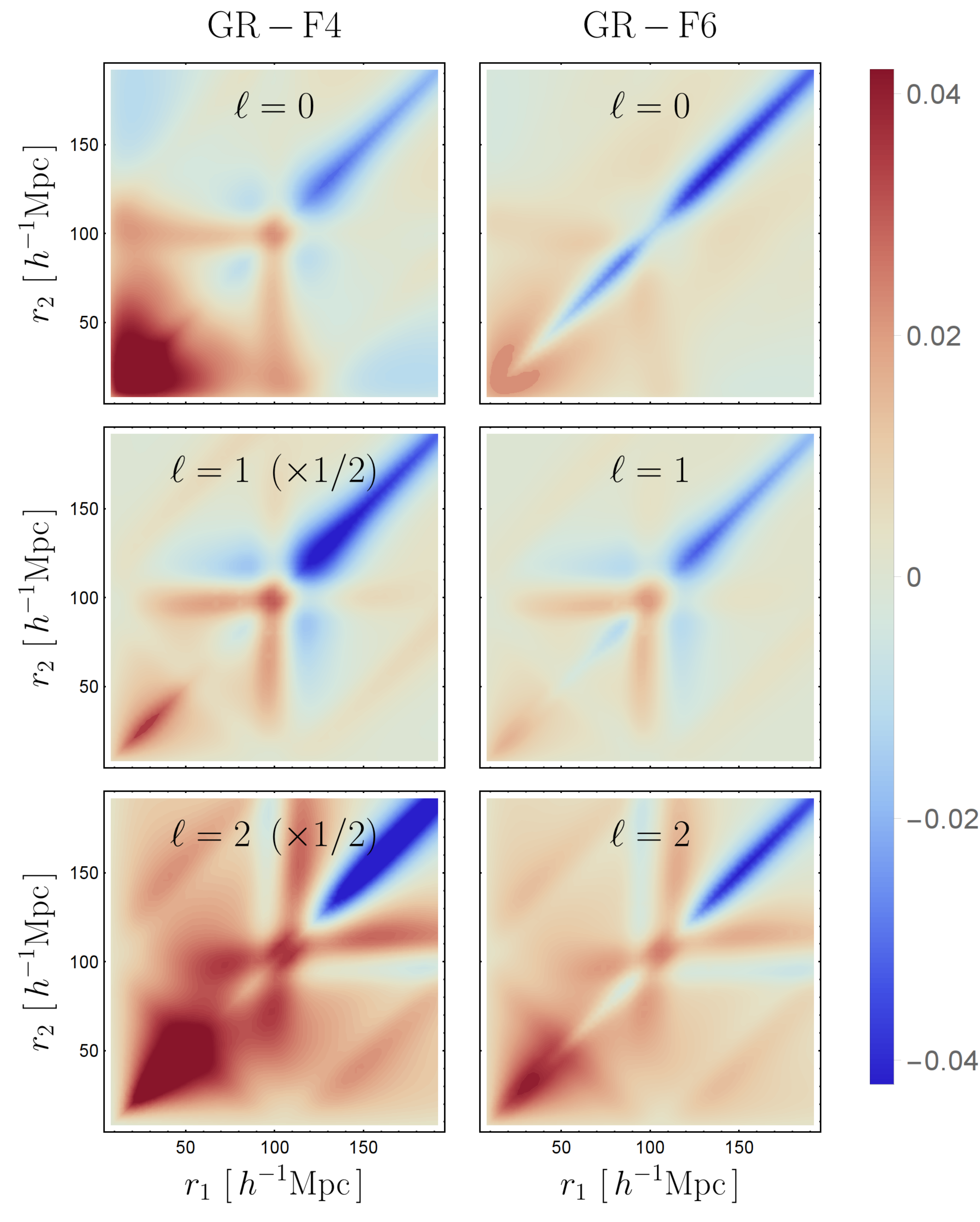}		
	\caption{Difference between GR and F4 models (left panel) and GR and F6 models (right panel) for the 3PCF of halos. The bias parameters that feed our Eq.~\eqref{pstcyc3PCF} are extracted from a halo catalogue of the \texttt{ELEPHANT} simulations with masses in the range $12.65< \log_{10}\big( M_h/h^{-1} M_\odot \big) < 13$.
	\label{fig:halos}}
	\end{center}
\end{figure*}

\begin{figure*}
	\begin{center}
	\includegraphics[width=7 in]{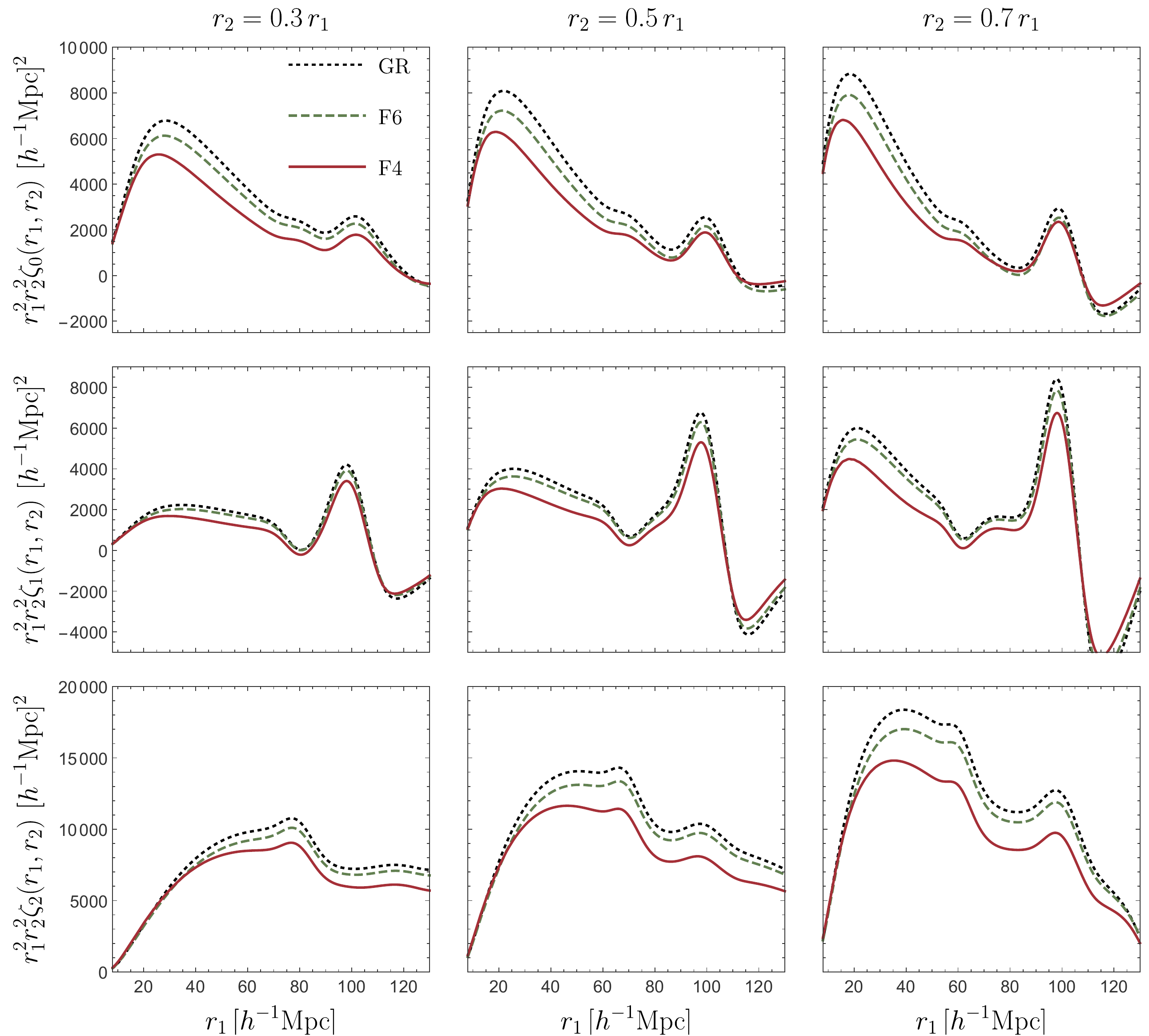}		
	\caption{1 dimensional plots for the 3PCF at directions $r_2=0.3 \,r_1$,  $r_2=0.5\, r_1$ and  $r_2=0.7 \,r_1$ (from left to right) and for multipoles $\ell=0,1,2$ (from top to bottom). We show the models GR (dotted black), F6 (dashed green) and F4 (red) at redshift $z=0.5$.
	\label{fig:halos1D}}
	\end{center}
\end{figure*}

The aim of this section is to present results obtained indirectly from simulated data. To that extent, we use bias parameters obtained in a previous work \cite{Aviles:2020wme} from halo catalogues of the Extended LEnsing PHysics using ANalaytic ray Tracing \texttt{ELEPHANT} $N$-body simulations, that were performed with a modified version of the \verb|RAMSES| code,   
the \verb|ECOSMOG| module \citep{1475-7516-2012-01-051,Bose:2016wms}. The cosmological parameters are fixed to $\Omega_m = 0.281$, $h=0.697$, $n_s=0.971$, $\Omega_b=0.046$ and $\sigma_8=0.848$, while 3 instances of the HS $n=1$ $f(R)$ model were simulated, corresponding to three variations of $|\bar{f}_{R_0}|=\{10^{-6},10^{-5}, 10^{-4}\}$ and as standadar referred as F6, F5, and F4. The simulations span a cubic volume of $V_{box}=(1024 \,\text{Mpc} \,h^{-1})^3$, with $1024^3$ dark matter particles.
We consider halos within the mass range $12.65< \log_{10}\big( M_h/h^{-1} M_\odot \big) < 13$, identified using the publicly available code \verb|ROCKSTAR| \citep{2013ApJ...762..109B}. These halos are named as halo catalog 2 in \cite{Aviles:2020wme}, or $z=0.5$ \verb|halos2| in table 1 of that paper. 
The bias parameters are given by
\begin{align}
 \text{GR}:&\quad \{b_1,b_2\}=\{1.655,-1 \},  \\
 \text{F6}:&\quad \{b_1,b_2\}=\{1.616,0.26 \}, \\
 \text{F5}:&\quad \{b_1,b_2\}=\{1.441,0.25 \}, 
\end{align}
while tidal bias is fixed by co-evolution \cite{Saito:2014qha}, $b_{s^2} = -\frac{2}{7}(b_1-1)$. Despite this expression is only valid for EdS evolution, it has proven to be accurate up to the sensitivity of the \verb|ELEPHANT| simulations \cite{Aviles:2020wme}.

In Fig.~\ref{fig:halos} we show the differences between GR and F4 models (left column) and between GR and F6 (right panels) for multipoles $\ell=0,1,2,3$. We do not plot the relative differences since the 3PCF crosses the zero several times. However, to have a sense of the differences in the signal for distinct gravity models, in Fig.~\ref{fig:halos1D} we plot 1-dimensional curves from arbitrary chosen directions in the plane $r_1$-$r_2$. From left to right, these are $r_2=0.3 \,r_1$,  $r_2=0.5\, r_1$ and  $r_2=0.7 \,r_1$, and for multipoles $\ell=0,1,2$ from top to bottom. We notice that the differences are indeed considerable and up to about 20\% in some regions, mainly due to that in MG the attractive extra fifth force tend to relax the CDM - baryons system more rapidly and the large scale bias goes towards unity faster than in GR.

\end{subsection}

\begin{section}{Conclusions}\label{subsec:concl}

In the study of the Large Scale Structure, higher order statistics will probe useful, not only to complement the analysis of two point statistics but also to unveil possible signals which are inherently associated to n-point correlations with $n>2$, such as consistency conditions (e.g. \cite{Kehagias:2013yd, Peloso:2013zw, Goldstein:2022hgr}), parity (e.g. \cite{Philcox:2022hkh,Hou:2022wfj}), particular shapes of primordial non-Gaussianities (e.g. \cite{Chen:2010xka}), to mention some. In the context of an initial Gaussian random field under the evolution of gravity, the non-linear gravitational interactions imprint non-trivial structures at all levels in the tower of n-point correlation functions of the matter distribution. In particular, constraining this structure in the context of three point correlations should be achievable with the stage IV galaxy surveys, as already discussed in the context of the DESI in \cite{Alam:2020jdv}. With this motivation in mind, we develop a theoretical framework for the three point correlation function (3PCF) of tracers in modified gravity, exemplified by two models with representative screening mechanisms: the Hu-Sawicki $f(R)$ \cite{Hu:2007nk} and the nDGP\cite{Dvali:2000hr} models. The final 3PCF result is expressed in a Legendre/Szapudi basis \cite{Szapudi:2004gg}, whose coefficients, given by Eq.~\eqref{pstcyc3PCF}, help not only to visualize deviations from GR (see for example \cite{SosaNunez:2020rpe}) but also to directly calculate the signal using estimators that scale as the two-point estimators with the number of data points \cite{Slepian:2015qza, Philcox:2021bwo}. 
We use standard perturbation theory with Effective Field Theory ingredients and a consistent biasing model to build up the tree-level MG bispectrum in the multipole basis, which are then directly map to the coefficients of the 3PCF in the same basis.

One would think that the Szapudi decomposition of Eq.~\eqref{zeta} is only meaningful if it rapidly converges to the full 3PCF. If that is the case, then a few multipoles will suffice, since the 3PCF should be smooth when galaxy fields are smeared over some reasonable scale. However, there is no proof of this convergence in the literature, or even a lack of studies on how much cosmological information the higher order multipoles contain. Actually, as it is apparent from Fig.~\ref{fig:postc3pcflm}, the amplitude of the multipoles do not seem to decay with $\ell$, with the warning  that our analysis only includes up to $\ell=8$. In the case of a slower convergence, one would need higher multipoles to recover the whole 3PCF from Eq.~\eqref{zeta}. However, it is important to point out that the 3PCF multipoles represent a summary statistics by themselves, which in turn can be used to constrain or rule out compelling cosmological theories. Moreover, the multipole expansion approach is somehow attractive because one can efficiently extract the signal from the data using only pair-counting algorithms, greatly reducing the prohibiting computational times of higher order estimators. As a final thought on this expansion, focusing on the different bias terms for each multipole (as in Fig.~\ref{fig:postc3pcflm}) allows for a deeper understanding of the imprinted gravitational structure in the 3PCF, where particular attention can be taken to the nearby region of the BAO scale for one or both of the triangle sides, and also around the subdiagonals or superdiagonals.

In terms of the relative differences between General Relativity and modify gravity, we appreciate a complex structure away from the diagonal in the 3PCF multipoles, as also described in \cite{Alam:2020jdv}. One may think of removing a fiducial LCDM signal to make more visible any GR deviation, in a similar spirit of the non-wiggle signal removal in the power spectrum. This may be useful, since we believe many of the other deviations from the canonical $\Lambda$CDM model that one could imagine would mostly sit along the diagonal or have more homogeneous signals. For example, a preliminary study of effect on the 3PCF from DESI's fiber allocation using the simulated data of \cite{Pinol:2016opt}, shows a very homogeneous difference in the monopole with short scale modifications along the diagonal in all the multipoles.

Natural extensions to this work are the inclusion of RSD, calculate 1-loop corrections, or even a further exploration of the consistency condition between the 3PCF and 2PCF in these MG models. Finally, it is worth stressing that some of the ideas presented here are shared by studies of the 3PCF multipole expansion in the presence of massive Neutrinos or scale-dependent primordial non-Gaussianities.

\end{section}

\acknowledgments

We would like to thank Baojiu Li and Zachary Slepian for useful discussions, and the Instituto Avanzado de Cosmolog\'ia A. C. for organising events where some of this work was done. A.~A. and G.N. acknowledge partial support by CONACyT Ciencia de Fronteras No. 102958.  A.~A. is also supported by CONACyT Ciencia de Frontera grant No.~319359 and CONACyT grant 283151, and acknowledges computational resources at DiRAC Data Centric system at Durham University. G.~N. also thanks the financial support of DAIP-UG and the computational resources of the DCI-UG DataLab.

\appendix

\begin{widetext}

\begin{section}{Curvature bispectrum and 3PCF multipoles}\label{app:Bispectrum}

In this appendix we derive the precyclic bispectrum in Eqs.~(\ref{BpcSplitcomp1})-(\ref{BpcSplitcomp6}) that are not shown in Eqs.(\ref{Bpclincomplete}).
That is we compute the multipoles of $B_{pc}^{\gamma_*}$, $B_{pc}^{\gamma_*\gamma}$ and $B_{pc}^{\gamma_*\gamma'}$, and their
Fourier transform that leads to the $Y_\ell$ functions.
As discussed in the main text, the contributions coming from terms $\gamma_* \Delta F_2$ are small, and are neglected in this work.
This can be introduced straightforwardly if needed, requiring the calculation of additional 2-dimensional integrals.
Hence, in this appendix we approximate $F_2 = F_{2,\text{LS}}$.

\vspace{1cm}

\noindent \underline{$ B_{\gamma_*}$}:
The $\gamma_*$ biasing component in the precyclic bispectrum is 
\begin{align}
 B_{pc,\gamma_*}(k_1,k_2,x)&=  (k_1^2 + k_2^2  + k_3^2) P_L(k_1)P_L(k_2) 2 F_2(\vk_1,\vk_2)  \nonumber\\
 &=  \Big[4 F_2(k_1,k_2,x)  (k_1^2 + k_2^2) + 4 xF_2(k_1,k_2,x) k_1 k_2 \Big] P_L(k_1)P_L(k_2),  
\end{align}
which upon integration against Eq.~\eqref{BexpcPlpc} gives the multipoles
\begin{align}
B_{pc,\ell=0}^{\gamma_*}(k_1,k_2) &=  \frac{70+12\mA^\text{LS}}{21}  (k_1^2 + k_2^2) P_L(k_1)P_L(k_2), \label{Bgsell0}\\
B_{pc,\ell=1}^{\gamma_*}(k_1,k_2) &= 
\frac{392 + 12 \mA^\text{LS}}{35}  k_1 k_2   P_L(k_1)P_L(k_2)   +  4\left(\frac{k_1^3}{k_2} + \frac{k_2^3}{k_1}\right) P_L(k_1)P_L(k_2), \label{Bgsell1}\\
B_{pc,\ell=2}^{\gamma_*}(k_1,k_2) &=\frac{56 - 12 \mA^\text{LS}}{21} (k_1^2 + k_2^2)   P_L(k_1)P_L(k_2), \label{Bgsell2}\\
B_{pc,\ell=3}^{\gamma_*}(k_1,k_2) &= \frac{28- 12 \mA^\text{LS}}{35}  k_1 k_2  P_L(k_1)P_L(k_2). \label{Bgsell3}
\end{align}
Their Fourier transforms give
\begin{align}
\zeta^{\gamma_*}_{pc,\ell=0}(r_1,r_2) &= \frac{70+12\mA^\text{LS}}{21} Y_0(r_1,r_2), \label{zgsY0}\\
\zeta^{\gamma_*}_{pc,\ell=1}(r_1,r_2) &= - \frac{392 + 12 \mA^\text{LS}}{35} Y_1(r_1,r_2), \\
\zeta^{\gamma_*}_{pc,\ell=2}(r_1,r_2)&= \frac{56 - 12 \mA^\text{LS}}{21} Y_2(r_1,r_2), \\
\zeta^{\gamma_*}_{pc,\ell=3}(r_1,r_2) &= -\frac{28- 12 \mA^\text{LS}}{35}  Y_3(r_1,r_2).\label{zgsY3}
\end{align}

\vspace{1cm}

\noindent \underline{$ B^{\gamma_* \gamma}$}:
The $\gamma_* \gamma$ biasing component in the precyclic bispectrum is 
$B_{pc,\ell}^{\gamma_* \gamma} = (k_1^2 + k_2^2) P_L(k_1)P_L(k_2)$, hence the multipoles are
\begin{align}\label{Bgsgell}
 B_{pc,\ell}^{\gamma_* \gamma}(k_1,k_2) = (k_1^2 + k_2^2) P_L(k_1)P_L(k_2) \delta_{\ell 0}
\end{align}
yielding a biasing $\gamma_* \gamma$ contribution to the 3PCF
\begin{align}
\zeta_{\gamma_* \gamma}^{\ell}(r_1,r_2)  =   Y_0(r_1,r_2) \delta_{\ell 0}
\end{align}

\vspace{1cm}

\noindent \underline{$ B_{\gamma_* \gamma'}$}:
The $\gamma_* \gamma'$ component in the precyclic bispectrum is 
$B_{pc,\ell}^{\gamma_* \gamma'} = \frac{2}{3}\p_2(x) (k_1^2 + k_2^2) P_L(k_1)P_L(k_2)$, hence the multipoles are
\begin{align} \label{Bgsgpell}
 B_{pc,\ell}^{\gamma_* \gamma'}(k_1,k_2) = \frac{2}{3} (k_1^2 + k_2^2) P_L(k_1)P_L(k_2) \delta_{\ell 2}
\end{align}
yielding a biasing $\gamma_* \gamma$ contribution to the 3PCF
\begin{align}
\zeta_{\gamma_* \gamma}^{\ell}(r_1,r_2)  =   \frac{2}{3} Y_2(r_1,r_2) \delta_{\ell 2}
\end{align}

\vspace{1cm}

\noindent \underline{$ B_{\gamma_*^2}$}

% \begin{align}
%  B_{pc,\gamma_*^2}(k_1,k_2,x) &=  (k_1^2 k_2^2 + k_2^2 k_3^2 + k_3^2 k_1^2) P_L(k_1)P_L(k_2) 2 F_2(k_1,k_2,x) \\
%  &= 2 (3k_1^2 k_2^2 + k_1^4 + k_2^4  )  F_2(k_1,k_2,x) P_L(k_1)P_L(k_2)   \nonumber\\
%  &\quad +  4 ( k_1^3k_2 +k_1k_2^3)    x F_2(k_1,k_2,x) P_L(k_1)P_L(k_2) 
% \end{align}

\begin{align}
 B_{pc,\gamma_*^2}(k_1,k_2,x) &=  (k_1^2 k_2^2 + k_2^2 k_3^2 + k_3^2 k_1^2) P_L(k_1)P_L(k_2) 2 F_2(k_1,k_2,x) \nonumber\\
 &= \Big[ 2 (3k_1^2 k_2^2 + k_1^4 + k_2^4  )  F_2(k_1,k_2,x)  +  4 ( k_1^3k_2 +k_1k_2^3)    x F_2(k_1,k_2,x) \Big] P_L(k_1)P_L(k_2) 
\end{align}

\begin{align}
B_{pc,\gamma_*^2}^{\ell=0}(k_1,k_2) &=  \frac{14 + 2 \mA^\text{LS}}{7} \Big[ k_1^4+k_2^4 +  \frac{56 + 9 \mA^\text{LS}}{21 + 3 \mA^\text{LS}}  k_1^2k_2^2 \Big]
 P_L(k_1)P_L(k_2), \\
B_{pc,\gamma_*^2}^{\ell=1}(k_1,k_2) &= 
\Big[ \frac{k_1^5}{k_2} +\frac{k_2^5}{k_1}  +
 \frac{252 + 12 \mA^\text{LS}}{35}(k_1^3 k_2 + k_1 k_2^3  ) \Big]  P_L(k_1)P_L(k_2), \\
B_{pc,\gamma_*^2}^{\ell=2}(k_1,k_2) &=\frac{14 - 2 \mA^\text{LS}}{21} \Big[ k_1^4 +  k_2^4  
 +\frac{49 + 3 \mA^\text{LS}}{21 + 3 \mA^\text{LS}}k_1^2 k_2^2   \Big] P_L(k_1)P_L(k_2), \\
B_{pc,\gamma_*^2}^{\ell=3}(k_1,k_2) &= \frac{28 - 12 \mA^\text{LS}}{35}(k_1^3 k^2 +  k_1 k_2^3) P_L(k_1)P_L(k_2).
\end{align}

\vspace{1cm}

\noindent \underline{$ B_{\gamma_*^3}$}

\begin{align}
 B_{\gamma_*^3}(k_1,k_2,x) &=  k_1^2 k_2^2 k^2_3 P_L(k_1)P_L(k_2) 2 F_2(k_1,k_2,x) \\
 &=  \Big[ 2 F_2(k_1,k_2,x) (k_1^4 k_2^2 + k_1^2k_2^4)
 + 4 x F_2(k_1,k_2,x) k_1^3 k_2^3 \Big] P_L(k_1)P_L(k_2)
\end{align}

\begin{align}
B_{pc,\gamma_*^3}^{\ell=0}(k_1,k_2) &=  \frac{14 + 2 \mA^\text{LS}}{7} (k_1^4 k_2^2 +k_1^2 k^4_2) P_L(k_1)P_L(k_2), \\
B_{pc,\gamma_*^3}^{\ell=1}(k_1,k_2) &= \left[ k_1^5 k_2 +k_1 k^2_5 + \frac{182 + 6 \mA^\text{LS}}{35} k_1^3 k_2^3 \right] P_L(k_1)P_L(k_2), \\
B_{pc,\gamma_*^3}^{\ell=2}(k_1,k_2) &= \frac{14 - 2 \mA^\text{LS}}{7} (k_1^4 k_2^2 +k_1^2 k^2_4) P_L(k_1)P_L(k_2), \\
B_{pc,\gamma_*^3}^{\ell=3}(k_1,k_2) &= \frac{28 - 12 \mA^\text{LS}}{35} k_1^3 k^3_2 P_L(k_1)P_L(k_2).     
\end{align}

\end{section}

\begin{section}{$F_2$ kernel computation}\label{app:F2}

%\emph{The derivation of the $F_2$ kernel in \cite{Aviles:2018saf} first finds the LPT kernels and from it the SPT kernels. I have a different derivation that uses SPT from the beginning. Perhaps we can write it here.}

%\aa{Here I write the direct EPT deduction of F2 and G2 kernels.}

In this appendix we derive the second order SPT kernels using directly the fluid equations. Perhaps the easiest route to get them is to obtain first the Lagrangian Perturbation Theory kernels \cite{Aviles:2017aor} and then perform a map to the Eulerian frame, e.g.~\cite{Aviles:2018saf,Aviles:2020wme}. However, for being self-contained in this work we show a direct derivation. To do so, we follow closely appendix A of \cite{Aviles:2021que} which performs the same computation but for cosmologies in the presence of massive neutrinos (see also \cite{Aviles:2020cax}). The main difference here is that we have to keep track of the screening terms. 
The continuity and Euler equations are
\begin{align}
\partial_t \delta(\vx,t) + \frac{1}{a} \partial_i v^i &=  - \frac{1}{a}\partial_i (v^i  \delta), \label{FScontEqCS}\\
 \partial_t v^i(\vx,t) +  H v^i
+ \frac{1}{a}\partial^i \Phi   &=   - \frac{1}{a} v^j \partial_j v^i, \label{FSEulerCS}
\end{align}
where $v^i$ is the velocity field of the fluid and $\delta$ its overdensity. These equations must be closed with the Poisson Eq.~\eqref{poisson} and Klein-Gordon Eq.~\eqref{KG}.
We define the $\theta$ field as
\begin{equation}
    \theta(\vx,t) = \frac{\partial_i v^i} { a H f_0 },
\end{equation}
with the growth rate 
\begin{equation}
  f(k,t) =\frac{ d\ln D_+(k,t)}{ d \ln a(t)}, \qquad f_0 \equiv f(k\rightarrow 0),  
\end{equation}
and the linear growth function solving the differential equation 
\begin{equation}
    \ddot{D}_+(k,t)  +  2H \dot{D}_+ = A(k) D_+,
\end{equation}
with $A(k)$ given by Eq.~\eqref{defAk}. Notice this equation has two solutions so one has to choose initial conditions that pick out the fastest growing solution. In the case where at early times one recovers GR, as for example in DGP and HS theories, one uses EdS initial conditions.  

We define the non-linear part of the self-interaction term of the Klein-Gordon Eq.~\eqref{KG} as, $\delta \mI = \mI - M_1\varphi$,
\begin{equation}
 \delta \mI = \frac{1}{2} \ikk M_2(\vk_1,\vk_2) \varphi(\vk_1) \varphi(\vk_2) + \cdots=   \frac{1}{2}\ikk  \mK_{\delta I}^{(2)}(\vk_1,\vk_2)\delta^{(1)}(\vk_1)\delta^{(1)}(\vk_2) + \cdots,
\end{equation}
where in the second equality the kernels $\mK_{\delta I}^{(n)}$ serve us to expand $\delta \mI$ in terms of linear density fields instead of the complete, non-linear scalar field $\varphi$.
With this, we can rewrite Eq.~\eqref{KG} in Fourier space as
\begin{align} \label{KG-F}
    \frac{1}{2}\varphi(\vk) = \frac{4\pi G \bar{\rho}_m}{3 \Pi(k)}\delta(\vk) - \frac{1}{6 \Pi(k)} \delta \mI(\vk),
\end{align}
with
\begin{equation}
    \Pi(k) = \frac{1}{6 a^2 \beta^2} \big( k^2 + m^2 a^2 \big).
\end{equation}
Using the Poisson Eq.~\eqref{poisson} and Eq.~\eqref{KG-F}, the fluid equations in Fourier space become\footnote{We use the shorthand notations
\begin{align}
 \underset{\vk_{1\cdots n}= \vk}{\int} = \int \Dk{k_1}\cdots\Dk{k_n} (2\pi)^3 \dD(\vk_{1\cdots n} - \vk), \qquad \vk_{1\cdots n} = \vk_1 + \cdots + \vk_n.   
\end{align}
}
\begin{align}
\frac{1}{H}  \frac{\partial\delta(\vk)}{\partial t} - f_0 \theta(\vk) &=  f_0  \ikk \alpha(\vk_1,\vk_2) \theta(\vk_1) \delta(\vk_2), \label{FScontEq}\\
 \frac{1}{H}  \frac{\partial f_0 \theta(\vk)}{\partial t} + \left( 2+ \frac{\dot{H}}{H^2}\right) f_0 \theta(\vk)
&- \frac{A(k)}{H^{2}} \delta(\vk) + \frac{k^2/a^2}{6 \Pi(k) H^2}\delta \mI(\vk)  =   f_0^2 \ikk \beta(\vk_1,\vk_2) \theta(\vk_1) \theta(\vk_2), \label{FSEulerEq}
\end{align}
with %$A(k)$ given by Eq.~\eqref{defAk} and, 
\begin{equation}
 \alpha(\vk_1,\vk_2) = 1+\frac{\vk_{1}\cdot\vk_2}{k_1^2},  \qquad  \beta(\vk_1,\vk_2) = \frac{k_{12}^2(\vk_1\cdot\vk_2)}{2 k_1^2 k_2^2}. 
\end{equation}

To linear order we obtain
\begin{align}
 \delta^{(1)}(\vk,t) = D_+(\vk,t)\delta^{(1)}(\vk,t_0),\qquad 
 \theta^{(1)}(\vk,t) = 
% -H^{-1} \dot{D}_+(\vk,t)\delta(\vk,t_0) = 
 \delta^{(1)}(\vk,t)
\end{align}
%with
%\begin{align}
% (\T - A(k)) D_+(\vk,t) = 0  
%\end{align}
and
\begin{equation}
    \varphi^{(1)}(\vk,t) = \frac{2 A_0}{3 \Pi(k)} \delta^{(1)}(\vk,t),
\end{equation}
with $A_0(t) = A(k\rightarrow 0,t) = 4\pi G \bar{\rho}_m  = 3 \Omega_m H^2/2$. Then, 
\begin{equation}\label{S2def}
S_2(\vk_1,\vk_2) \equiv \frac{k^2/a^2}{6\Pi(k)}  \mK_{\delta I}^{(2)}(\vk_1,\vk_2) =  \left(\frac{2 A_0}{3 }\right)^2\frac{ M_2(\vk_1,\vk_2) k^2/a^2}{6 \Pi(k)\Pi(k_1)\Pi(k_2)}.
\end{equation}
%
%and the definition of the rescaled velocity divergence $\theta = -i k_i v^i / (aHf_0)$ in terms of the peculiar velocity $v^i$. % in eq.~\eqref{thetacb}. 

We introduce the SPT kernels through
\begin{align}
 \delta^{(n)} (\vk,t) &= \underset{\vk_{1\cdots n}= \vk}{\int} F_n(\vk_1,\cdots,\vk_n;t) \delta^{(1)}(\vk_1,t) \cdots \delta^{(1)}(\vk_n,t), \nonumber\\
 \theta^{(n)} (\vk,t) &= \underset{\vk_{1\cdots n}= \vk}{\int} G_n(\vk_1,\cdots,\vk_n;t) \delta^{(1)}(\vk_1,t) \cdots \delta^{(1)}(\vk_n,t).
\end{align}
Hence, at first order, 
\begin{align}
 F_1(\vk) &=1, \qquad \text{and} \qquad
 G_1(\vk) = \frac{f(k)}{f_0},
\end{align}

To second order, the fluid equations are
\begin{align}
 H^{-1}\frac{\partial\delta^{(2)}(\vk)}{\partial t} - f_0\theta^{(2)}(\vk) &=  f_0 \ikk \alpha(\vk_1,\vk_2) \theta^{(1)}(\vk_1) \delta^{(1)}(\vk_2) \nonumber \\
 &=\frac{1}{2}\ikk \big[\alpha(\vk_1,\vk_2)f(k_1) + \alpha(\vk_2,\vk_1)f(k_2) \big]\delta^{(1)}(\vk_1) \delta^{(1)}(\vk_2) \label{ContFS2},
\end{align}
\begin{align} 
 H^{-1}\frac{\partial f_0 \theta^{(2)}(\vk)}{\partial t} + f_0 \left( 2 + \frac{\dot{H}}{H^2}\right) \theta^{(2)}(\vk)
& - \frac{A(k)}{H^{2}} \delta^{(2)}(\vk)  + \frac{1}{2} \ikk S_2(\vk_1,\vk_2) \delta^{(1)}(\vk_1) \delta^{(1)}(\vk_2) \nonumber\\ & =   f_0^2 \ikk \beta(\vk_1,\vk_2) \theta^{(1)}(\vk_1) \theta^{(1)}(\vk_2), \nonumber\\ 
&=    \ikk \beta(\vk_1,\vk_2) f(k_1)f(k_2) \delta^{(1)}(\vk_1) \delta^{(1)}(\vk_2) \label{EulerFS2},
\end{align}
where we have used $\theta^{(1)}(\vk) = (f(k)/f_0)\delta^{(1)}(\vk)$, and inside the integral of the rhs of eq.~\eqref{ContFS2} we have symmetrized over. % (Warning: one should not do this symmetrization if one wants to compute $F_3$ and $G_3$ with the obtained $F_2$ and $G_2$).

The second order overdensity and velocity fields are
\begin{align}
 \delta^{(2)} (\vk) &= \underset{\vk_{12}= \vk}{\int} F_2(\vk_1,\vk_2) D_+(\vk_1,t)D_+(\vk_2,t)\delta^{(1)}(\vk_1,t_0)\delta^{(1)}(\vk_2,t_0), \nonumber\\
 \theta^{(2)} (\vk) &= \underset{\vk_{12}= \vk}{\int} G_2(\vk_1,\vk_2) D_+(\vk_1,t)D_+(\vk_2,t)\delta^{(1)}(\vk_1,t_0)\delta^{(1)}(\vk_2,t_0).
\end{align}
%with $\delta_0(\vk) = \delta^{(1)}(\vk,t_0)$. 
Inserting these expressions into eqs.~\eqref{ContFS2} and \eqref{EulerFS2},
\begin{align}
 &\frac{1 }{H D_1D_2} \frac{d \,\,}{d t}(F_2 D_1D_2) - f_0 G_2 = 
            \frac{1}{2} (\alpha_{12}f_1+\alpha_{21}f_2), \\
 &\frac{1 }{H D_1D_2} \frac{d \,\,}{d t}(f_0 G_2 D_1D_2)     + \left( 2 + \frac{\dot{H}}{H^2}\right) f_0 G_2
   -\frac{A(k)}{H^2}F_2   = - \frac{S_2}{2} + \beta_{12} f_1 f_2,
\end{align}
with $f_{1,2}=f(k_{1,2})$, $D_{1,2}=D_+(\vk_{1,2},t)$, $\alpha_{12}=\alpha(\vk_1,\vk_2)$, $\alpha_{21}=\alpha(\vk_2,\vk_1)$ and $\beta_{12}=\beta(\vk_1,\vk_2)$. 
We rewrite the above equations as
\begin{align}
&\frac{1}{H}  \frac{d F_2}{d t} + F_2(f_1+f_2) - f_0 G_2 =  \frac{1}{2} (\alpha_{12}f_1+\alpha_{21}f_2), \label{Ceq3}\\
&\frac{1}{H}  \frac{d f_0 G_2}{d t} + f_0 G_2(f_1+f_2) + \left( 2 + \frac{\dot{H}}{H^2}\right) f_0 G_2
   -\frac{A(k)}{H^2}F_2   = - \frac{S_2}{2} + \beta_{12} f_1 f_2. \label{Eeq3}
\end{align}
% where in the second equality of eq.~\eqref{Ceq3} we have symmetrized $\alpha_{12}f_1 \rightarrow \frac{1}{2} (\alpha_{12}f_1+\alpha_{21}f_2)$, which is valid inside the $\int_{\vk_{12}=\vk}$ integral. 

Taking the time derivative of eq.~\eqref{Ceq3} and using eq.~\eqref{Eeq3} we obtain a second order equation for $F_2$,
\begin{align}
& \frac{1}{H^2}\ddot{F}_2 + \frac{2}{H}(1+f_1+f_2)\dot{F}_2 + \left[ \frac{1}{H}(\dot{f}_1 + \dot{f}_2) 
 + (f_1+f_2)\left(f_1+f_2 + 2 + \frac{\dot{H}}{H^2} \right) - \frac{A(k)}{H^2}\right] F_2 = \nonumber\\
& \qquad \frac{1}{2 H} (\alpha_{12}\dot{f}_1 + \alpha_{21}\dot{f}_2) +  \frac{1}{2 } (\alpha_{12}f_1 + \alpha_{21}f_2)\left(f_1+f_2 + 2 + \frac{\dot{H}}{H^2} \right)
+ \beta_{12}f_1f_2  - \frac{S_2}{2},  \label{ddotF}
\end{align}
Now, the growth rate $f(k)$ evolves as
\begin{equation}\label{dotfeq}
  \dot{f} = \frac{A(k)}{H} - H \left(2+\frac{\dot{H}}{H^2} \right)f  - H f^2.
\end{equation}
Substituting for $\dot{f}_1$ and $\dot{f}_2$ in eq.~\eqref{ddotF},
\begin{align}\label{ddotF2}
& \frac{1}{H^2} \ddot{F}_2 + \frac{2}{H}(1+f_1+f_2)\dot{F}_2 + \left[ 2f_1 f_2 + \frac{A(k_1)+A(k_2) - A(k)}{H^2} \right] F_2
 =  \nonumber\\ &\qquad \frac{1}{2}\alpha_{12}\frac{A(k_1)}{H^2} + \frac{1}{2}\alpha_{21}\frac{A(k_2)}{H^2}
  + \frac{1}{2}f_1f_2(\alpha_{12}+\alpha_{21}) + \beta_{12}f_1f_2  - \frac{S_2}{2}.
\end{align}
Now, let us define a second order growth function as
\begin{equation}
 D^{(2)}(\vk_1,\vk_2,t) \equiv D_{12} \equiv 2 D_1 D_2 F_2 - \chi_{12},
\end{equation}
and
\begin{align}
\chi_{12} &\equiv \alpha_{12}+\alpha_{21} - \gamma_{12}, \qquad \text{with} \qquad
\gamma_{12} \equiv 1 - \frac{(\vk_1\cdot\vk_2)^2}{k_1^2 k_2^2}. 
\end{align}
We will now find a differential equation for $D_{12}$.
First, the second order $F_2$ kernel is
\begin{equation}\label{F2ansatz}
 F_2 = \frac{D_{12}}{2D_1D_2} + \frac{1}{2}\chi_{12}.
\end{equation}

Now, taking time derivatives of the above equation, 
\begin{equation}
\frac{1}{H}\dot{F}_2 = \frac{1}{2D_1D_2}\left( \frac{1}{H}\dot{D}_{12} - D_{12}(f_1+f_2) \right),
\end{equation}

\begin{align}
  \frac{1}{H^2}\ddot{F}_2 &= \frac{1}{2D_1D_2}\Bigg[ \frac{1}{H^2}\ddot{D}_{12} - \frac{2}{H}(f_1+f_2)\dot{D}_{12} \nonumber\\
  &\quad \qquad +D_{12}\left( 2 (f_1^2 + f_2^2 + f_1f_2 +f_1+f_2)  - \frac{1}{H^2}(A(k_1)+ A(k_2))\right)\Bigg],
\end{align}
where we used eq.~\eqref{dotfeq} and $\ddot{D} + 2H \dot{D} = A(k)D$.
Substituting the above equations into eq.~\eqref{ddotF2},
\begin{align}
 \ddot{D}_{12} + 2H\dot{D}_{12} - A(k)D_{12} &= \Bigg[ A(k)  
 + (A(k)-A(k_2)) \frac{\vk_1 \cdot \vk_2}{k^2_1}
 + (A(k)-A(k_1)) \frac{\vk_1 \cdot \vk_2}{k^2_2} \nonumber\\ 
 &\quad \qquad - (A(k_1)+A(k_2) - A(k))\frac{(\vk_1\cdot\vk_2)^2}{k_1^2k_2^2}   - S_2\Bigg] D_1 D_2.
\end{align}
Hence, using eq.~\eqref{F2ansatz} we obtain
\begin{align}
F_2(\vk_1,\vk_2) &= \frac{1}{2} + \frac{3}{14}\mA + \left( \frac{1}{2} - \frac{3}{14}\mB  \right)   \frac{(\vk_1\cdot\vk_2)^2}{k_1^2 k_2^2}
        + \frac{\vk_1\cdot\vk_2}{2 k_1k_2} \left(\frac{k_2}{k_1} + \frac{k_1}{k_2} \right), \label{F2_kernelapp}
\end{align}
and from eq.~\eqref{Ceq3},
\begin{align}
G_2(\vk_1,\vk_2) &= \frac{3\mA(f_1+f_2) + 3 \dot{\mA}/H }{14 f_0} +
\left(\frac{f_1+f_2}{2 f_0} - \frac{3\mB(f_1+f_2) + 3 \dot{\mB}/H }{14 f_0}\right) \frac{(\vk_1\cdot\vk_2)^2}{k_1^2 k_2^2} \nonumber\\
&\quad + \frac{\vk_1\cdot\vk_2}{2 k_1k_2} \left( \frac{f_2}{f_0}\frac{k_2}{k_1} + \frac{f_1}{f_0}\frac{k_1}{k_2} \right), \label{G2_kernelapp}
\end{align}
with $\mA$ and $\mB$ given by
\begin{equation} \label{AandBdef2}
 \mA(\vk_1,\vk_2,t) = \frac{7 D^{(2)}_{\mA}(\vk_1,\vk_2,t)}{3 D_{+}(k_1,t)D_{+}(k_2,t)}, 
 \qquad \mB(\vk_1,\vk_2,t) = \frac{7 D^{(2)}_{\mB}(\vk_1,\vk_2,t)}{3 D_{+}(k_1,t)D_{+}(k_2,t)},
\end{equation}
with second order growth functions $D^{(2)}_{\mA}$ and $D^{(2)}_{\mB}$ the solutions to
\begin{align}
D^{(2)}_{\mA} = \big(\T - A(k)\big)^{-1}\Big[A(k) &+ (A(k)-A(k_1) ) \frac{\vk_1\cdot \vk_2}{k_2^2}  +(A(k)-A(k_2) ) \frac{\vk_1\cdot \vk_2}{k_1^2} - S_2(\vk_1,\vk_2)\Big]  D_{+}(k_1)D_{+}(k_2), \label{DAeveq2} \\
D^{(2)}_{\mB} = \big(\T - A(k)\big)^{-1} \Big[A(k_1) &+ A(k_2) - A(k) \Big]  D_{+}(k_1)D_{+}(k_2), \label{DBeveq2}
\end{align}
with $k=|\vk_1 + \vk_2|$, which are Eqs.~\eqref{DAeveq} and \eqref{DBeveq}

\end{section}

\end{widetext}

 \bibliographystyle{JHEP}  % Use the "unsrtnat" BibTeX style for formatting the Bibliography
 \bibliography{refs.bib}

\providecommand{\href}[2]{#2}\begingroup\raggedright\begin{thebibliography}{10}

\bibitem{Verde:2019ivm}
L.~Verde, T.~Treu and A.~G. Riess, \emph{{Tensions between the Early and the
  Late Universe}},
  \href{http://dx.doi.org/10.1038/s41550-019-0902-0}{\emph{Nature Astron.} {\bf
  3} (7, 2019) 891}, [\href{http://arxiv.org/abs/1907.10625}{{\tt
  1907.10625}}].

\bibitem{2016arXiv161100036D}
{DESI Collaboration}, A.~{Aghamousa}, J.~{Aguilar}, S.~{Ahlen}, S.~{Alam},
  L.~E. {Allen} et~al., \emph{{The DESI Experiment Part I: Science,Targeting,
  and Survey Design}}, {\emph{arXiv e-prints} (Oct., 2016) arXiv:1611.00036},
  [\href{http://arxiv.org/abs/1611.00036}{{\tt 1611.00036}}].

\bibitem{2016arXiv161100037D}
{DESI Collaboration}, A.~{Aghamousa}, J.~{Aguilar}, S.~{Ahlen}, S.~{Alam},
  L.~E. {Allen} et~al., \emph{{The DESI Experiment Part II: Instrument
  Design}}, {\emph{arXiv e-prints} (Oct., 2016) arXiv:1611.00037},
  [\href{http://arxiv.org/abs/1611.00037}{{\tt 1611.00037}}].

\bibitem{2022arXiv220510939A}
{DESI Collaboration}, B.~{Abareshi}, J.~{Aguilar}, S.~{Ahlen}, S.~{Alam}, D.~M.
  {Alexander} et~al., \emph{{Overview of the Instrumentation for the Dark
  Energy Spectroscopic Instrument}}, {\emph{arXiv e-prints} (May, 2022)
  arXiv:2205.10939}, [\href{http://arxiv.org/abs/2205.10939}{{\tt
  2205.10939}}].

\bibitem{EUCLID:2011zbd}
{\scshape EUCLID} collaboration, R.~Laureijs et~al., \emph{{Euclid Definition
  Study Report}},  \href{http://arxiv.org/abs/1110.3193}{{\tt 1110.3193}}.

\bibitem{LSST:2008ijt}
{\scshape LSST} collaboration, v.~Ivezi\'c et~al., \emph{{LSST: from Science
  Drivers to Reference Design and Anticipated Data Products}},
  \href{http://dx.doi.org/10.3847/1538-4357/ab042c}{\emph{Astrophys. J.} {\bf
  873} (2019) 111}, [\href{http://arxiv.org/abs/0805.2366}{{\tt 0805.2366}}].

\bibitem{Lovelock:1971yv}
D.~Lovelock, \emph{{The Einstein tensor and its generalizations}},
  \href{http://dx.doi.org/10.1063/1.1665613}{\emph{J. Math. Phys.} {\bf 12}
  (1971) 498--501}.

\bibitem{Lovelock:1972vz}
D.~Lovelock, \emph{{The four-dimensionality of space and the einstein tensor}},
  \href{http://dx.doi.org/10.1063/1.1666069}{\emph{J. Math. Phys.} {\bf 13}
  (1972) 874--876}.

\bibitem{Will:2014kxa}
C.~M. Will, \emph{{The Confrontation between General Relativity and
  Experiment}}, \href{http://dx.doi.org/10.12942/lrr-2014-4}{\emph{Living Rev.
  Rel.} {\bf 17} (2014) 4}, [\href{http://arxiv.org/abs/1403.7377}{{\tt
  1403.7377}}].

\bibitem{Khoury:2010xi}
J.~Khoury, \emph{{Theories of Dark Energy with Screening Mechanisms}},
  \href{http://arxiv.org/abs/1011.5909}{{\tt 1011.5909}}.

\bibitem{Khoury:2003aq}
J.~Khoury and A.~Weltman, \emph{{Chameleon fields: Awaiting surprises for tests
  of gravity in space}},
  \href{http://dx.doi.org/10.1103/PhysRevLett.93.171104}{\emph{Phys. Rev.
  Lett.} {\bf 93} (2004) 171104},
  [\href{http://arxiv.org/abs/astro-ph/0309300}{{\tt astro-ph/0309300}}].

\bibitem{Hinterbichler:2010es}
K.~Hinterbichler and J.~Khoury, \emph{{Symmetron Fields: Screening Long-Range
  Forces Through Local Symmetry Restoration}},
  \href{http://dx.doi.org/10.1103/PhysRevLett.104.231301}{\emph{Phys. Rev.
  Lett.} {\bf 104} (2010) 231301}, [\href{http://arxiv.org/abs/1001.4525}{{\tt
  1001.4525}}].

\bibitem{Vainshtein:1972sx}
A.~I. Vainshtein, \emph{{To the problem of nonvanishing gravitation mass}},
  \href{http://dx.doi.org/10.1016/0370-2693(72)90147-5}{\emph{Phys. Lett.} {\bf
  39B} (1972) 393--394}.

\bibitem{DeFelice:2010aj}
A.~De~Felice and S.~Tsujikawa, \emph{{f(R) theories}},
  \href{http://dx.doi.org/10.12942/lrr-2010-3}{\emph{Living Rev. Rel.} {\bf 13}
  (2010) 3}, [\href{http://arxiv.org/abs/1002.4928}{{\tt 1002.4928}}].

\bibitem{Hu:2007nk}
W.~Hu and I.~Sawicki, \emph{{ Models of f(R) Cosmic Acceleration that Evade
  Solar-System Tests}},
  \href{http://dx.doi.org/10.1103/PhysRevD.76.064004}{\emph{Phys. Rev. D} {\bf
  76} (2007) 064004}, [\href{http://arxiv.org/abs/0705.1158}{{\tt 0705.1158}}].

\bibitem{Dvali:2000hr}
G.~R. Dvali, G.~Gabadadze and M.~Porrati, \emph{{4-D gravity on a brane in 5-D
  Minkowski space}},
  \href{http://dx.doi.org/10.1016/S0370-2693(00)00669-9}{\emph{Phys. Lett. B}
  {\bf 485} (2000) 208--214}, [\href{http://arxiv.org/abs/hep-th/0005016}{{\tt
  hep-th/0005016}}].

\bibitem{Slepian:2015qza}
Z.~Slepian and D.~J. Eisenstein, \emph{{Computing the three-point correlation
  function of galaxies in $\mathcal {O}(N^2)$ time}},
  \href{http://dx.doi.org/10.1093/mnras/stv2119}{\emph{Mon. Not. Roy. Astron.
  Soc.} {\bf 454} (2015) 4142--4158},
  [\href{http://arxiv.org/abs/1506.02040}{{\tt 1506.02040}}].

\bibitem{Slepian:2015qwa}
Z.~Slepian and D.~J. Eisenstein, \emph{{Accelerating the two-point and
  three-point galaxy correlation functions using Fourier transforms}},
  \href{http://dx.doi.org/10.1093/mnrasl/slv133}{\emph{Mon. Not. Roy. Astron.
  Soc.} {\bf 455} (2016) L31--L35},
  [\href{http://arxiv.org/abs/1506.04746}{{\tt 1506.04746}}].

\bibitem{Slepian:2017lpm}
Z.~Slepian and D.~J. Eisenstein, \emph{{A practical computational method for
  the anisotropic redshift-space three-point correlation function}},
  \href{http://dx.doi.org/10.1093/mnras/sty1063}{\emph{Mon. Not. Roy. Astron.
  Soc.} {\bf 478} (2018) 1468--1483},
  [\href{http://arxiv.org/abs/1709.10150}{{\tt 1709.10150}}].

\bibitem{Slepian:2016kfz}
Z.~Slepian et~al., \emph{{Detection of baryon acoustic oscillation features in
  the large-scale three-point correlation function of SDSS BOSS DR12 CMASS
  galaxies}}, \href{http://dx.doi.org/10.1093/mnras/stx488}{\emph{Mon. Not.
  Roy. Astron. Soc.} {\bf 469} (2017) 1738--1751},
  [\href{http://arxiv.org/abs/1607.06097}{{\tt 1607.06097}}].

\bibitem{Slepian:2016nfb}
Z.~Slepian et~al., \emph{{Constraining the baryon–dark matter relative
  velocity with the large-scale three-point correlation function of the SDSS
  BOSS DR12 CMASS galaxies}},
  \href{http://dx.doi.org/10.1093/mnras/stx2723}{\emph{Mon. Not. Roy. Astron.
  Soc.} {\bf 474} (2018) 2109--2115},
  [\href{http://arxiv.org/abs/1607.06098}{{\tt 1607.06098}}].

\bibitem{Alam:2020jdv}
S.~Alam et~al., \emph{{Towards testing the theory of gravity with DESI: summary
  statistics, model predictions and future simulation requirements}},
  \href{http://dx.doi.org/10.1088/1475-7516/2021/11/050}{\emph{JCAP} {\bf 11}
  (2021) 050}, [\href{http://arxiv.org/abs/2011.05771}{{\tt 2011.05771}}].

\bibitem{Cautun:2017tkc}
M.~Cautun, E.~Paillas, Y.-C. Cai, S.~Bose, J.~Armijo, B.~Li et~al., \emph{{The
  Santiago\textendash{}Harvard\textendash{}Edinburgh\textendash{}Durham void
  comparison \textendash{} I. SHEDding light on chameleon gravity tests}},
  \href{http://dx.doi.org/10.1093/mnras/sty463}{\emph{Mon. Not. Roy. Astron.
  Soc.} {\bf 476} (2018) 3195--3217},
  [\href{http://arxiv.org/abs/1710.01730}{{\tt 1710.01730}}].

\bibitem{Hellwing:2017pmj}
W.~A. Hellwing, K.~Koyama, B.~Bose and G.-B. Zhao, \emph{{Revealing modified
  gravity signals in matter and halo hierarchical clustering}},
  \href{http://dx.doi.org/10.1103/PhysRevD.96.023515}{\emph{Phys. Rev.} {\bf
  D96} (2017) 023515}, [\href{http://arxiv.org/abs/1703.03395}{{\tt
  1703.03395}}].

\bibitem{Li:2011vk}
B.~Li, G.-B. Zhao, R.~Teyssier and K.~Koyama, \emph{{ECOSMOG: An Efficient Code
  for Simulating Modified Gravity}},
  \href{http://dx.doi.org/10.1088/1475-7516/2012/01/051}{\emph{JCAP} {\bf 1201}
  (2012) 051}, [\href{http://arxiv.org/abs/1110.1379}{{\tt 1110.1379}}].

\bibitem{Bose:2016wms}
S.~Bose, B.~Li, A.~Barreira, J.-h. He, W.~A. Hellwing, K.~Koyama et~al.,
  \emph{{Speeding up $N$-body simulations of modified gravity: Chameleon
  screening models}},
  \href{http://dx.doi.org/10.1088/1475-7516/2017/02/050}{\emph{JCAP} {\bf 1702}
  (2017) 050}, [\href{http://arxiv.org/abs/1611.09375}{{\tt 1611.09375}}].

\bibitem{Li:2013nua}
B.~Li, G.-B. Zhao and K.~Koyama, \emph{{Exploring Vainshtein mechanism on
  adaptively refined meshes}},
  \href{http://dx.doi.org/10.1088/1475-7516/2013/05/023}{\emph{JCAP} {\bf 1305}
  (2013) 023}, [\href{http://arxiv.org/abs/1303.0008}{{\tt 1303.0008}}].

\bibitem{Barreira:2015xvp}
A.~Barreira, S.~Bose and B.~Li, \emph{{Speeding up N-body simulations of
  modified gravity: Vainshtein screening models}},
  \href{http://dx.doi.org/10.1088/1475-7516/2015/12/059}{\emph{JCAP} {\bf 1512}
  (2015) 059}, [\href{http://arxiv.org/abs/1511.08200}{{\tt 1511.08200}}].

\bibitem{Koyama:2009me}
K.~Koyama, A.~Taruya and T.~Hiramatsu, \emph{{Non-linear Evolution of Matter
  Power Spectrum in Modified Theory of Gravity}},
  \href{http://dx.doi.org/10.1103/PhysRevD.79.123512}{\emph{Phys. Rev.} {\bf
  D79} (2009) 123512}, [\href{http://arxiv.org/abs/0902.0618}{{\tt
  0902.0618}}].

\bibitem{Brax:2013fna}
P.~Brax and P.~Valageas, \emph{{Impact on the power spectrum of Screening in
  Modified Gravity Scenarios}},
  \href{http://dx.doi.org/10.1103/PhysRevD.88.023527}{\emph{Phys. Rev.} {\bf
  D88} (2013) 023527}, [\href{http://arxiv.org/abs/1305.5647}{{\tt
  1305.5647}}].

\bibitem{Taruya:2014faa}
A.~Taruya, T.~Nishimichi, F.~Bernardeau, T.~Hiramatsu and K.~Koyama,
  \emph{{Regularized cosmological power spectrum and correlation function in
  modified gravity models}},
  \href{http://dx.doi.org/10.1103/PhysRevD.90.123515}{\emph{Phys. Rev.} {\bf
  D90} (2014) 123515}, [\href{http://arxiv.org/abs/1408.4232}{{\tt
  1408.4232}}].

\bibitem{Bellini:2015oua}
E.~Bellini and M.~Zumalacarregui, \emph{{Nonlinear evolution of the baryon
  acoustic oscillation scale in alternative theories of gravity}},
  \href{http://dx.doi.org/10.1103/PhysRevD.92.063522}{\emph{Phys. Rev.} {\bf
  D92} (2015) 063522}, [\href{http://arxiv.org/abs/1505.03839}{{\tt
  1505.03839}}].

\bibitem{Fasiello:2017bot}
M.~Fasiello and Z.~Vlah, \emph{{Screening in perturbative approaches to LSS}},
  \href{http://arxiv.org/abs/1704.07552}{{\tt 1704.07552}}.

\bibitem{Aviles:2017aor}
A.~Aviles and J.~L. Cervantes-Cota, \emph{{Lagrangian perturbation theory for
  modified gravity}},
  \href{http://dx.doi.org/10.1103/PhysRevD.96.123526}{\emph{Phys. Rev.} {\bf
  D96} (2017) 123526}, [\href{http://arxiv.org/abs/1705.10719}{{\tt
  1705.10719}}].

\bibitem{Bose:2018orj}
B.~Bose, K.~Koyama, M.~Lewandowski, F.~Vernizzi and H.~A. Winther,
  \emph{{Towards Precision Constraints on Gravity with the Effective Field
  Theory of Large-Scale Structure}},
  \href{http://dx.doi.org/10.1088/1475-7516/2018/04/063}{\emph{JCAP} {\bf 1804}
  (2018) 063}, [\href{http://arxiv.org/abs/1802.01566}{{\tt 1802.01566}}].

\bibitem{Aviles:2020wme}
A.~Aviles, G.~Valogiannis, M.~A. Rodriguez-Meza, J.~L. Cervantes-Cota, B.~Li
  and R.~Bean, \emph{{Redshift space power spectrum beyond Einstein-de Sitter
  kernels}}, \href{http://dx.doi.org/10.1088/1475-7516/2021/04/039}{\emph{JCAP}
  {\bf 04} (2021) 039}, [\href{http://arxiv.org/abs/2012.05077}{{\tt
  2012.05077}}].

\bibitem{Cataneo:2018cic}
M.~Cataneo, L.~Lombriser, C.~Heymans, A.~Mead, A.~Barreira, S.~Bose et~al.,
  \emph{{On the road to per-cent accuracy: nonlinear reaction of the matter
  power spectrum to dark energy and modified gravity}},
  \href{http://arxiv.org/abs/1812.05594}{{\tt 1812.05594}}.

\bibitem{Taruya:2013quf}
A.~Taruya, K.~Koyama, T.~Hiramatsu and A.~Oka, \emph{{Beyond consistency test
  of gravity with redshift-space distortions at quasilinear scales}},
  \href{http://dx.doi.org/10.1103/PhysRevD.89.043509}{\emph{Phys. Rev.} {\bf
  D89} (2014) 043509}, [\href{http://arxiv.org/abs/1309.6783}{{\tt
  1309.6783}}].

\bibitem{Bose:2016qun}
B.~Bose and K.~Koyama, \emph{{A Perturbative Approach to the Redshift Space
  Power Spectrum: Beyond the Standard Model}},
  \href{http://dx.doi.org/10.1088/1475-7516/2016/08/032}{\emph{JCAP} {\bf 1608}
  (2016) 032}, [\href{http://arxiv.org/abs/1606.02520}{{\tt 1606.02520}}].

\bibitem{Aviles:2018saf}
A.~Aviles, M.~A. Rodriguez-Meza, J.~De-Santiago and J.~L. Cervantes-Cota,
  \emph{{Nonlinear evolution of initially biased tracers in modified gravity}},
  \href{http://dx.doi.org/10.1088/1475-7516/2018/11/013}{\emph{JCAP} {\bf 1811}
  (2018) 013}, [\href{http://arxiv.org/abs/1809.07713}{{\tt 1809.07713}}].

\bibitem{Valogiannis:2019xed}
G.~Valogiannis and R.~Bean, \emph{{Convolution Lagrangian Perturbation Theory
  for biased tracers beyond general relativity}},
  \href{http://arxiv.org/abs/1901.03763}{{\tt 1901.03763}}.

\bibitem{Hirano:2018uar}
S.~Hirano, T.~Kobayashi, H.~Tashiro and S.~Yokoyama, \emph{{Matter bispectrum
  beyond Horndeski theories}},
  \href{http://dx.doi.org/10.1103/PhysRevD.97.103517}{\emph{Phys. Rev.} {\bf
  D97} (2018) 103517}, [\href{http://arxiv.org/abs/1801.07885}{{\tt
  1801.07885}}].

\bibitem{Bose:2018zpk}
B.~Bose and A.~Taruya, \emph{{The one-loop matter bispectrum as a probe of
  gravity and dark energy}},  \href{http://arxiv.org/abs/1808.01120}{{\tt
  1808.01120}}.

\bibitem{Bose:2019wuz}
B.~Bose, J.~Byun, F.~Lacasa, A.~Moradinezhad~Dizgah and L.~Lombriser,
  \emph{{Modelling the non-linear bispectrum in modified gravity}},
  \href{http://arxiv.org/abs/1909.02504}{{\tt 1909.02504}}.

\bibitem{Slepian:2016weg}
Z.~Slepian and D.~J. Eisenstein, \emph{{Modelling the large-scale
  redshift-space 3-point correlation function of galaxies}},
  \href{http://dx.doi.org/10.1093/mnras/stx490}{\emph{Mon. Not. Roy. Astron.
  Soc.} {\bf 469} (2017) 2059--2076},
  [\href{http://arxiv.org/abs/1607.03109}{{\tt 1607.03109}}].

\bibitem{Desjacques:2016bnm}
V.~Desjacques, D.~Jeong and F.~Schmidt, \emph{{Large-Scale Galaxy Bias}},
  \href{http://dx.doi.org/10.1016/j.physrep.2017.12.002}{\emph{Phys. Rept.}
  {\bf 733} (2018) 1--193}, [\href{http://arxiv.org/abs/1611.09787}{{\tt
  1611.09787}}].

\bibitem{Clifton:2011jh}
T.~Clifton, P.~G. Ferreira, A.~Padilla and C.~Skordis, \emph{{Modified Gravity
  and Cosmology}},
  \href{http://dx.doi.org/10.1016/j.physrep.2012.01.001}{\emph{Phys. Rept.}
  {\bf 513} (2012) 1--189}, [\href{http://arxiv.org/abs/1106.2476}{{\tt
  1106.2476}}].

\bibitem{Horndeski:1974wa}
G.~W. Horndeski, \emph{{Second-order scalar-tensor field equations in a
  four-dimensional space}},
  \href{http://dx.doi.org/10.1007/BF01807638}{\emph{Int. J. Theor. Phys.} {\bf
  10} (1974) 363--384}.

\bibitem{Langlois:2015cwa}
D.~Langlois and K.~Noui, \emph{{Degenerate higher derivative theories beyond
  Horndeski: evading the Ostrogradski instability}},
  \href{http://dx.doi.org/10.1088/1475-7516/2016/02/034}{\emph{JCAP} {\bf 02}
  (2016) 034}, [\href{http://arxiv.org/abs/1510.06930}{{\tt 1510.06930}}].

\bibitem{Crisostomi:2016czh}
M.~Crisostomi, K.~Koyama and G.~Tasinato, \emph{{Extended Scalar-Tensor
  Theories of Gravity}},
  \href{http://dx.doi.org/10.1088/1475-7516/2016/04/044}{\emph{JCAP} {\bf 04}
  (2016) 044}, [\href{http://arxiv.org/abs/1602.03119}{{\tt 1602.03119}}].

\bibitem{Joyce:2014kja}
A.~Joyce, B.~Jain, J.~Khoury and M.~Trodden, \emph{{Beyond the Cosmological
  Standard Model}},
  \href{http://dx.doi.org/10.1016/j.physrep.2014.12.002}{\emph{Phys. Rept.}
  {\bf 568} (2015) 1--98}, [\href{http://arxiv.org/abs/1407.0059}{{\tt
  1407.0059}}].

\bibitem{Babichev:2009ee}
E.~Babichev, C.~Deffayet and R.~Ziour, \emph{{k-Mouflage gravity}},
  \href{http://dx.doi.org/10.1142/S0218271809016107}{\emph{Int. J. Mod. Phys.}
  {\bf D18} (2009) 2147--2154}, [\href{http://arxiv.org/abs/0905.2943}{{\tt
  0905.2943}}].

\bibitem{Nicolis:2008in}
A.~Nicolis, R.~Rattazzi and E.~Trincherini, \emph{{The Galileon as a local
  modification of gravity}},
  \href{http://dx.doi.org/10.1103/PhysRevD.79.064036}{\emph{Phys. Rev.} {\bf
  D79} (2009) 064036}, [\href{http://arxiv.org/abs/0811.2197}{{\tt
  0811.2197}}].

\bibitem{Matsubara:2015ipa}
T.~Matsubara, \emph{{Recursive Solutions of Lagrangian Perturbation Theory}},
  \href{http://dx.doi.org/10.1103/PhysRevD.92.023534}{\emph{Phys. Rev.} {\bf
  D92} (2015) 023534}, [\href{http://arxiv.org/abs/1505.01481}{{\tt
  1505.01481}}].

\bibitem{Schmidt:2009sv}
F.~Schmidt, \emph{{Cosmological Simulations of Normal-Branch Braneworld
  Gravity}}, \href{http://dx.doi.org/10.1103/PhysRevD.80.123003}{\emph{Phys.
  Rev.} {\bf D80} (2009) 123003}, [\href{http://arxiv.org/abs/0910.0235}{{\tt
  0910.0235}}].

\bibitem{Pietroni:2005pv}
M.~Pietroni, \emph{{Dark energy condensation}},
  \href{http://dx.doi.org/10.1103/PhysRevD.72.043535}{\emph{Phys. Rev. D} {\bf
  72} (2005) 043535}, [\href{http://arxiv.org/abs/astro-ph/0505615}{{\tt
  astro-ph/0505615}}].

\bibitem{Olive:2007aj}
K.~A. Olive and M.~Pospelov, \emph{{Environmental dependence of masses and
  coupling constants}},
  \href{http://dx.doi.org/10.1103/PhysRevD.77.043524}{\emph{Phys. Rev. D} {\bf
  77} (2008) 043524}, [\href{http://arxiv.org/abs/0709.3825}{{\tt 0709.3825}}].

\bibitem{Aviles:2018qot}
A.~Aviles, J.~L. Cervantes-Cota and D.~F. Mota, \emph{{Screenings in Modified
  Gravity: a perturbative approach}},
  \href{http://dx.doi.org/10.1051/0004-6361/201834383}{\emph{Astron.
  Astrophys.} {\bf 622} (2019) A62},
  [\href{http://arxiv.org/abs/1810.02652}{{\tt 1810.02652}}].

\bibitem{Lewis:1999bs}
A.~Lewis, A.~Challinor and A.~Lasenby, \emph{{Efficient computation of CMB
  anisotropies in closed FRW models}},
  \href{http://dx.doi.org/10.1086/309179}{\emph{Astrophys. J.} {\bf 538} (2000)
  473--476}, [\href{http://arxiv.org/abs/astro-ph/9911177}{{\tt
  astro-ph/9911177}}].

\bibitem{Hojjati:2011ix}
A.~Hojjati, L.~Pogosian and G.-B. Zhao, \emph{{Testing gravity with CAMB and
  CosmoMC}}, \href{http://dx.doi.org/10.1088/1475-7516/2011/08/005}{\emph{JCAP}
  {\bf 1108} (2011) 005}, [\href{http://arxiv.org/abs/1106.4543}{{\tt
  1106.4543}}].

\bibitem{Blas:2011rf}
D.~Blas, J.~Lesgourgues and T.~Tram, \emph{{The Cosmic Linear Anisotropy
  Solving System (CLASS) II: Approximation schemes}},
  \href{http://dx.doi.org/10.1088/1475-7516/2011/07/034}{\emph{JCAP} {\bf 1107}
  (2011) 034}, [\href{http://arxiv.org/abs/1104.2933}{{\tt 1104.2933}}].

\bibitem{Zumalacarregui:2016pph}
M.~Zumalac\'arregui, E.~Bellini, I.~Sawicki, J.~Lesgourgues and P.~G. Ferreira,
  \emph{{hi\_class: Horndeski in the Cosmic Linear Anisotropy Solving System}},
  \href{http://dx.doi.org/10.1088/1475-7516/2017/08/019}{\emph{JCAP} {\bf 1708}
  (2017) 019}, [\href{http://arxiv.org/abs/1605.06102}{{\tt 1605.06102}}].

\bibitem{McDonald:2009dh}
P.~McDonald and A.~Roy, \emph{{Clustering of dark matter tracers: generalizing
  bias for the coming era of precision LSS}},
  \href{http://dx.doi.org/10.1088/1475-7516/2009/08/020}{\emph{JCAP} {\bf 0908}
  (2009) 020}, [\href{http://arxiv.org/abs/0902.0991}{{\tt 0902.0991}}].

\bibitem{Assassi:2014fva}
V.~Assassi, D.~Baumann, D.~Green and M.~Zaldarriaga, \emph{{Renormalized Halo
  Bias}}, \href{http://dx.doi.org/10.1088/1475-7516/2014/08/056}{\emph{JCAP}
  {\bf 1408} (2014) 056}, [\href{http://arxiv.org/abs/1402.5916}{{\tt
  1402.5916}}].

\bibitem{Hui:2007zh}
L.~Hui and K.~P. Parfrey, \emph{{The Evolution of Bias: Generalized}},
  \href{http://dx.doi.org/10.1103/PhysRevD.77.043527}{\emph{Phys. Rev.} {\bf
  D77} (2008) 043527}, [\href{http://arxiv.org/abs/0712.1162}{{\tt
  0712.1162}}].

\bibitem{Parfrey:2010uy}
K.~Parfrey, L.~Hui and R.~K. Sheth, \emph{{Scale-dependent halo bias from
  scale-dependent growth}},
  \href{http://dx.doi.org/10.1103/PhysRevD.83.063511}{\emph{Phys. Rev.} {\bf
  D83} (2011) 063511}, [\href{http://arxiv.org/abs/1012.1335}{{\tt
  1012.1335}}].

\bibitem{Kaiser:1984sw}
N.~Kaiser, \emph{{On the Spatial correlations of Abell clusters}},
  \href{http://dx.doi.org/10.1086/184341}{\emph{Astrophys. J.} {\bf 284} (1984)
  L9--L12}.

\bibitem{Bardeen:1985tr}
J.~M. Bardeen, J.~R. Bond, N.~Kaiser and A.~S. Szalay, \emph{{The Statistics of
  Peaks of Gaussian Random Fields}},
  \href{http://dx.doi.org/10.1086/164143}{\emph{Astrophys. J.} {\bf 304} (1986)
  15--61}.

\bibitem{Desjacques:2010gz}
V.~Desjacques, M.~Crocce, R.~Scoccimarro and R.~K. Sheth, \emph{{Modeling
  scale-dependent bias on the baryonic acoustic scale with the statistics of
  peaks of Gaussian random fields}},
  \href{http://dx.doi.org/10.1103/PhysRevD.82.103529}{\emph{Phys. Rev.} {\bf
  D82} (2010) 103529}, [\href{http://arxiv.org/abs/1009.3449}{{\tt
  1009.3449}}].

\bibitem{Lazeyras:2015giz}
T.~Lazeyras, M.~Musso and V.~Desjacques, \emph{{Lagrangian bias of generic
  large-scale structure tracers}},
  \href{http://dx.doi.org/10.1103/PhysRevD.93.063007}{\emph{Phys. Rev.} {\bf
  D93} (2016) 063007}, [\href{http://arxiv.org/abs/1512.05283}{{\tt
  1512.05283}}].

\bibitem{Schmidt:2012ys}
F.~Schmidt, D.~Jeong and V.~Desjacques, \emph{{Peak-Background Split,
  Renormalization, and Galaxy Clustering}},
  \href{http://dx.doi.org/10.1103/PhysRevD.88.023515}{\emph{Phys. Rev.} {\bf
  D88} (2013) 023515}, [\href{http://arxiv.org/abs/1212.0868}{{\tt
  1212.0868}}].

\bibitem{Aviles:2018thp}
A.~Aviles, \emph{Renormalization of lagrangian bias via spectral parameters},
  \href{http://dx.doi.org/10.1103/PhysRevD.98.083541}{\emph{Phys. Rev.} {\bf
  D98} (2018) 083541}, [\href{http://arxiv.org/abs/1805.05304}{{\tt
  1805.05304}}].

\bibitem{Baumann:2010tm}
D.~Baumann, A.~Nicolis, L.~Senatore and M.~Zaldarriaga, \emph{{Cosmological
  Non-Linearities as an Effective Fluid}},
  \href{http://dx.doi.org/10.1088/1475-7516/2012/07/051}{\emph{JCAP} {\bf 07}
  (2012) 051}, [\href{http://arxiv.org/abs/1004.2488}{{\tt 1004.2488}}].

\bibitem{Pietroni:2011iz}
M.~Pietroni, G.~Mangano, N.~Saviano and M.~Viel, \emph{{Coarse-Grained
  Cosmological Perturbation Theory}},
  \href{http://dx.doi.org/10.1088/1475-7516/2012/01/019}{\emph{JCAP} {\bf 01}
  (2012) 019}, [\href{http://arxiv.org/abs/1108.5203}{{\tt 1108.5203}}].

\bibitem{Porto:2013qua}
R.~A. Porto, L.~Senatore and M.~Zaldarriaga, \emph{{The Lagrangian-space
  Effective Field Theory of Large Scale Structures}},
  \href{http://dx.doi.org/10.1088/1475-7516/2014/05/022}{\emph{JCAP} {\bf 05}
  (2014) 022}, [\href{http://arxiv.org/abs/1311.2168}{{\tt 1311.2168}}].

\bibitem{Vlah:2015sea}
Z.~Vlah, M.~White and A.~Aviles, \emph{{A Lagrangian effective field theory}},
  \href{http://dx.doi.org/10.1088/1475-7516/2015/09/014}{\emph{JCAP} {\bf 09}
  (2015) 014}, [\href{http://arxiv.org/abs/1506.05264}{{\tt 1506.05264}}].

\bibitem{Ivanov:2021kcd}
M.~M. Ivanov, O.~H.~E. Philcox, T.~Nishimichi, M.~Simonovi\'c, M.~Takada and
  M.~Zaldarriaga, \emph{{Precision analysis of the redshift-space galaxy
  bispectrum}},
  \href{http://dx.doi.org/10.1103/PhysRevD.105.063512}{\emph{Phys. Rev. D} {\bf
  105} (2022) 063512}, [\href{http://arxiv.org/abs/2110.10161}{{\tt
  2110.10161}}].

\bibitem{Philcox:2021kcw}
O.~H.~E. Philcox and M.~M. Ivanov, \emph{{BOSS DR12 full-shape cosmology:
  \ensuremath{\Lambda}CDM constraints from the large-scale galaxy power
  spectrum and bispectrum monopole}},
  \href{http://dx.doi.org/10.1103/PhysRevD.105.043517}{\emph{Phys. Rev. D} {\bf
  105} (2022) 043517}, [\href{http://arxiv.org/abs/2112.04515}{{\tt
  2112.04515}}].

\bibitem{Philcox:2021bwo}
O.~H.~E. Philcox, Z.~Slepian, J.~Hou, C.~Warner, R.~N. Cahn and D.~J.
  Eisenstein, \emph{{encore: an O (Ng2) estimator for galaxy N-point
  correlation functions}},
  \href{http://dx.doi.org/10.1093/mnras/stab3025}{\emph{Mon. Not. Roy. Astron.
  Soc.} {\bf 509} (2021) 2457--2481},
  [\href{http://arxiv.org/abs/2105.08722}{{\tt 2105.08722}}].

\bibitem{Philcox:2021hbm}
O.~H.~E. Philcox, J.~Hou and Z.~Slepian, \emph{{A First Detection of the
  Connected 4-Point Correlation Function of Galaxies Using the BOSS CMASS
  Sample}},  \href{http://arxiv.org/abs/2108.01670}{{\tt 2108.01670}}.

\bibitem{Szapudi:2004gg}
I.~Szapudi, \emph{{Three - point statistics from a new perspective}},
  \href{http://dx.doi.org/10.1086/420894}{\emph{Astrophys. J.} {\bf 605} (2004)
  L89}, [\href{http://arxiv.org/abs/astro-ph/0404476}{{\tt astro-ph/0404476}}].

\bibitem{Valogiannis:2019nfz}
G.~Valogiannis, R.~Bean and A.~Aviles, \emph{{An accurate perturbative approach
  to redshift space clustering of biased tracers in modified gravity}},
  \href{http://arxiv.org/abs/1909.05261}{{\tt 1909.05261}}.

\bibitem{Li:2011qda}
B.~Li and G.~Efstathiou, \emph{{An Extended Excursion Set Approach to Structure
  Formation in Chameleon Models}},
  \href{http://dx.doi.org/10.1111/j.1365-2966.2011.20404.x}{\emph{Mon. Not.
  Roy. Astron. Soc.} {\bf 421} (2012) 1431},
  [\href{http://arxiv.org/abs/1110.6440}{{\tt 1110.6440}}].

\bibitem{Kopp:2013lea}
M.~Kopp, S.~A. Appleby, I.~Achitouv and J.~Weller, \emph{{Spherical collapse
  and halo mass function in $f(R)$ theories}},
  \href{http://dx.doi.org/10.1103/PhysRevD.88.084015}{\emph{Phys. Rev. D} {\bf
  88} (2013) 084015}, [\href{http://arxiv.org/abs/1306.3233}{{\tt 1306.3233}}].

\bibitem{Slepian:2014dda}
Z.~Slepian and D.~Eisenstein, \emph{{On the signature of the baryon–dark
  matter relative velocity in the two- and three-point galaxy correlation
  functions}}, \href{http://dx.doi.org/10.1093/mnras/stu2627}{\emph{Mon. Not.
  Roy. Astron. Soc.} {\bf 448} (2015) 9--26},
  [\href{http://arxiv.org/abs/1411.4052}{{\tt 1411.4052}}].

\bibitem{Koyama:2013paa}
K.~Koyama, G.~Niz and G.~Tasinato, \emph{{Effective theory for the Vainshtein
  mechanism from the Horndeski action}},
  \href{http://dx.doi.org/10.1103/PhysRevD.88.021502}{\emph{Phys. Rev. D} {\bf
  88} (2013) 021502}, [\href{http://arxiv.org/abs/1305.0279}{{\tt 1305.0279}}].

\bibitem{1475-7516-2012-01-051}
B.~Li, G.-B. Zhao, R.~Teyssier and K.~Koyama, \emph{Ecosmog : an efficient code
  for simulating modified gravity}, {\emph{Journal of Cosmology and
  Astroparticle Physics} {\bf 2012} (2012) 051}.

\bibitem{2013ApJ...762..109B}
P.~S. {Behroozi}, R.~H. {Wechsler} and H.-Y. {Wu}, \emph{{The ROCKSTAR
  Phase-space Temporal Halo Finder and the Velocity Offsets of Cluster Cores}},
  \href{http://dx.doi.org/10.1088/0004-637X/762/2/109}{\emph{The Astrophysical
  Journal} {\bf 762} (Jan., 2013) 109},
  [\href{http://arxiv.org/abs/1110.4372}{{\tt 1110.4372}}].

\bibitem{Saito:2014qha}
S.~Saito, T.~Baldauf, Z.~Vlah, U.~Seljak, T.~Okumura and P.~McDonald,
  \emph{{Understanding higher-order nonlocal halo bias at large scales by
  combining the power spectrum with the bispectrum}},
  \href{http://dx.doi.org/10.1103/PhysRevD.90.123522}{\emph{Phys. Rev.} {\bf
  D90} (2014) 123522}, [\href{http://arxiv.org/abs/1405.1447}{{\tt
  1405.1447}}].

\bibitem{Kehagias:2013yd}
A.~Kehagias and A.~Riotto, \emph{{Symmetries and Consistency Relations in the
  Large Scale Structure of the Universe}},
  \href{http://dx.doi.org/10.1016/j.nuclphysb.2013.05.009}{\emph{Nucl. Phys. B}
  {\bf 873} (2013) 514--529}, [\href{http://arxiv.org/abs/1302.0130}{{\tt
  1302.0130}}].

\bibitem{Peloso:2013zw}
M.~Peloso and M.~Pietroni, \emph{{Galilean invariance and the consistency
  relation for the nonlinear squeezed bispectrum of large scale structure}},
  \href{http://dx.doi.org/10.1088/1475-7516/2013/05/031}{\emph{JCAP} {\bf 05}
  (2013) 031}, [\href{http://arxiv.org/abs/1302.0223}{{\tt 1302.0223}}].

\bibitem{Goldstein:2022hgr}
S.~Goldstein, A.~Esposito, O.~H.~E. Philcox, L.~Hui, J.~C. Hill, R.~Scoccimarro
  et~al., \emph{{Squeezing ${f_{NL}}$ out of the matter bispectrum with
  consistency relations}},  \href{http://arxiv.org/abs/2209.06228}{{\tt
  2209.06228}}.

\bibitem{Philcox:2022hkh}
O.~H.~E. Philcox, \emph{{Probing parity violation with the four-point
  correlation function of BOSS galaxies}},
  \href{http://dx.doi.org/10.1103/PhysRevD.106.063501}{\emph{Phys. Rev. D} {\bf
  106} (2022) 063501}, [\href{http://arxiv.org/abs/2206.04227}{{\tt
  2206.04227}}].

\bibitem{Hou:2022wfj}
J.~Hou, Z.~Slepian and R.~N. Cahn, \emph{{Measurement of Parity-Odd Modes in
  the Large-Scale 4-Point Correlation Function of SDSS BOSS DR12 CMASS and LOWZ
  Galaxies}},  \href{http://arxiv.org/abs/2206.03625}{{\tt 2206.03625}}.

\bibitem{Chen:2010xka}
X.~Chen, \emph{{Primordial Non-Gaussianities from Inflation Models}},
  \href{http://dx.doi.org/10.1155/2010/638979}{\emph{Adv. Astron.} {\bf 2010}
  (2010) 638979}, [\href{http://arxiv.org/abs/1002.1416}{{\tt 1002.1416}}].

\bibitem{SosaNunez:2020rpe}
F.~Sosa Nu\~nez and G.~Niz, \emph{{On the fast random sampling and other
  properties of the three point correlation function in galaxy surveys}},
  \href{http://dx.doi.org/10.1088/1475-7516/2020/12/021}{\emph{JCAP} {\bf 12}
  (2020) 021}, [\href{http://arxiv.org/abs/2006.05434}{{\tt 2006.05434}}].

\bibitem{Pinol:2016opt}
L.~Pinol, R.~N. Cahn, N.~Hand, U.~Seljak and M.~White, \emph{{Imprint of DESI
  fiber assignment on the anisotropic power spectrum of emission line
  galaxies}},
  \href{http://dx.doi.org/10.1088/1475-7516/2017/04/008}{\emph{JCAP} {\bf 04}
  (2017) 008}, [\href{http://arxiv.org/abs/1611.05007}{{\tt 1611.05007}}].

\bibitem{Aviles:2021que}
A.~Aviles, A.~Banerjee, G.~Niz and Z.~Slepian, \emph{{Clustering in massive
  neutrino cosmologies via Eulerian Perturbation Theory}},
  \href{http://dx.doi.org/10.1088/1475-7516/2021/11/028}{\emph{JCAP} {\bf 11}
  (2021) 028}, [\href{http://arxiv.org/abs/2106.13771}{{\tt 2106.13771}}].

\bibitem{Aviles:2020cax}
A.~Aviles and A.~Banerjee, \emph{{A Lagrangian Perturbation Theory in the
  presence of massive neutrinos}},
  \href{http://dx.doi.org/10.1088/1475-7516/2020/10/034}{\emph{JCAP} {\bf 10}
  (2020) 034}, [\href{http://arxiv.org/abs/2007.06508}{{\tt 2007.06508}}].

\end{thebibliography}\endgroup

\end{document}